\documentclass[12pt]{article}
\usepackage{subeqnarray}
\usepackage{cite}
\usepackage{epsf}

\newlength{\titlesep}
\newlength{\authorsep}
\makeatletter

\renewcommand{\thesection}{\Roman{section}}

\def\fnum@figure{FIG.~\thefigure}

\newcommand{\reffig}[1]{Fig.~\protect\ref{#1}}

\@addtoreset{equation}{section}
\renewcommand{\theequation}{\arabic{section}.\arabic{equation}}

\renewcommand{\abstract}{\if@twocolumn
  \section*{Abstract}
  \else
  \begin{center}
    {\bf Abstract\vspace{-.5em}\vspace{0pt}}
  \end{center}
  \fi}
\renewcommand{\endabstract}{\if@twocolumn\else\endquotation\fi}

\renewcommand{\appendix}{\par
  \setcounter{section}{0}
  \setcounter{subsection}{0}
  \renewcommand{\thesection}{Appendix~\Alph{section}}
  \renewcommand{\theequation}{(\Alph{section}.\arabic{equation})}}

\newcommand{\thismonth}{\ifcase\month\or
 January\or February\or March\or April\or May\or June\or
 July\or August\or September\or October\or November\or December\fi
 \space \number\year}

\newcommand{\preprintnumber}[1]
{{\bf\begin{flushright}%
\begin{tabular}{l}#1\end{tabular}
\end{flushright}}}

\newcommand{\refsec}[1]{Sec.~\protect\ref{#1}}

\makeatother

\newcommand{\ie}{{\it i.e.\/}}
\newcommand{\etal}{{\it et al.\/}}
\newcommand{\etc}{{\it etc.\/}}
\newcommand{\ol}[1]{\overline{#1}}
\newcommand{\wt}[1]{\widetilde{#1}}
\newcommand{\vev}[1]{\left\langle #1 \right\rangle}
\newcommand{\hc}{{\rm h.\,c.}\,}
\newcommand{\sign}{\mathop{\rm sign}\nolimits}
\newcommand{\br}{\mathop{\rm B}\nolimits}   
\newcommand{\gsim}%
{\mathrel{\mbox{\raisebox{-1.0ex}
    {$\stackrel{\displaystyle >}{\displaystyle \sim}$}}}}
\newcommand{\lsim}%
{\mathrel{\mbox{\raisebox{-1.0ex}
    {$\stackrel{\displaystyle <}{\displaystyle \sim}$}}}}

\newcommand{\tanb}{\ensuremath{\tan\beta}}
\newcommand{\bb}{$B^0$--$\ol{B}^0$}
\newcommand{\kk}{$K^0$--$\ol{K}^0$}
\newcommand{\ek}{\ensuremath{\epsilon_K}}
\newcommand{\bsg}{\ensuremath{b\rightarrow s\, \gamma}}
\newcommand{\brbsg}{\ensuremath{\br(\bsg)}}
\newcommand{\bsll}{\ensuremath{b\rightarrow s\, l^+\, l^-}}
\newcommand{\kpnn}{\ensuremath{K\rightarrow \pi\, \nu\, \ol{\nu}}}
\newcommand{\klpnn}{\ensuremath{K_L\rightarrow \pi^0\, \nu\, \ol{\nu}}}
\newcommand{\kppnn}{\ensuremath{K^+\rightarrow \pi^+\, \nu\, \ol{\nu}}}
\newcommand{\dmbd}{\ensuremath{\Delta M_{B_d}}}

\newcommand{\rxd}{\ensuremath{\dmbd\slash\dmbd^{\rm SM}}}
\newcommand{\rek}{\ensuremath{\ek\slash\ek^{\rm SM}}}
\newcommand{\rklpnn}{\ensuremath{\br(\klpnn)\slash\br(\klpnn)_{\rm SM}}}
\newcommand{\rkppnn}{\ensuremath{\br(\kppnn)\slash\br(\kppnn)_{\rm SM}}}
\newcommand{\amu}{\ensuremath{a_\mu^{\rm SUSY}}}
\newcommand{\Journal}[4]{{#1} {\bf #2}, {#4} {(#3)}}
\newcommand{\xxx}[1]{{\tt #1}}

\newcommand{\plb}{\sl Phys.~Lett.~{\bf B}}

\newcommand{\prd}{\sl Phys.~Rev.~{\bf D}}
\newcommand{\prl}{\sl Phys.~Rev.~Lett.}
\newcommand{\np}{\sl Nucl.~Phys.}
\newcommand{\npb}{\sl Nucl.~Phys.~{\bf B}}
\newcommand{\ptp}{\sl Prog.~Theor.~Phys.}
\newcommand{\ptps}{\sl Prog.~Theor.~Phys.~Suppl.}

\newcommand{\zpc}{\sl Z.~Phys.~{\bf C}}

\newcommand{\mpla}{\sl Mod.~Phys.~Lett.~{\bf A}}

\newcommand{\ibid}{\it ibid.}
\newcommand{\epj}{\it Eur.~Phys.~J.}
\newcommand{\jhp}{\it JHEP}

\begin{document}
\baselineskip 18pt

\begin{titlepage}
\preprintnumber{%
KEK-TH-611 \\
August 1999
}
\vspace*{\titlesep}
\begin{center}
{\LARGE\bf
Flavor changing neutral current processes and the muon anomalous
magnetic moment in the supergravity model
}
\\
\vspace*{\titlesep}
{\large
\footnote{
E-mail: toru.goto@kek.jp.
}%
Toru Goto,
\footnote{
E-mail: yasuhiro.okada@kek.jp.
}%
Yasuhiro Okada,
and
\footnote{
E-mail: yasuhiro.shimizu@kek.jp.
}%
Yasuhiro Shimizu
}
\\
\vspace*{\authorsep}
{\it
Theory Group, KEK, Tsukuba, Ibaraki, 305-0801 Japan
}
\end{center}
\vspace*{\titlesep}
\begin{abstract}
Flavor changing neutral current processes in $B$ and $K$ decays in the
supergravity model are revisited taking into account the recent progress 
of Higgs boson search experiments at LEP.
Possible deviations of \bb\ mixing, the CP violating \kk\ mixing
parameter \ek\ from the standard model predictions are reduced
significantly in a small \tanb\ region due to the constraints on the
SUSY parameter space imposed by the Higgs boson search.
With the present bound on the Higgs boson mass, the magnitude of the
\bb\ mixing amplitude can be enhanced up to 12\% for $\tanb=3$ in the
minimal supergravity.
If we relax the strict universality of SUSY breaking scalar masses at
the GUT scale, the deviation can be 25\%.
We also investigate SUSY contributions to the muon anomalous magnetic
moment in the supergravity model and show that there is a correlation
with the  \bsg\ amplitude.
The SUSY contribution to $a_\mu = \frac{(g-2)_\mu}{2}$ can be ($-30$ --
$+80)\times 10^{-10}$ for $\tan\beta=10$ and even larger for larger
$\tan\beta$.
The improved measurement of  the muon anomalous magnetic moment in the
on-going experiment at BNL will put a strong constraint on SUSY
parameter space, especially for a large \tanb\ region.
\end{abstract}

\end{titlepage}

\section{Introduction}

In order to look for physics beyond the standard model (SM), low energy
experiments can play an important role.
Various flavor changing neutral current (FCNC) processes in $B$ and $K$
meson decays and the muon anomalous magnetic moment could receive large
contributions from loop diagrams of new particles.
On-going experiments at KEK and SLAC $B$-factories and the muon storage
ring at BNL as well as future plans of $B$ and $K$ meson decay
experiments therefore may be able to reveal physics beyond the energy
scale which is currently accessible from direct particle search
experiments.

There have been extensive studies on FCNC effects in the supergravity
model.
In a supersymmetry (SUSY) model, the SUSY particles can give large
contributions to various FCNC processes such as
\brbsg\cite{GM-BG,BBMR,bsg}, \bb\ and \kk\ mixings,
\cite{bbbar,GM-BG,BBMR,GNO} and $\br(\kpnn)$ \cite{kpnn}. 
These contributions depend on the flavor mixing in the squark mass
matrix which is {\it a priori} independent of the flavor mixing in the
quark sector.
In fact, FCNC processes can put strong constraints on these mixing
parameters because arbitrary mixing parameters tend to give too large
effects on, for instance, the \kk\ mixing amplitude if the SUSY particles
exist below 1 TeV.
In the minimal supergravity model, these dangerous FCNCs are eliminated
by the assumption that all squarks have universal SUSY breaking mass
terms through the flavor-blind coupling of gravity interaction.
In the previous publication \cite{GOS} we have investigated the \bb\
mixing, the CP violating parameter of \kk\ mixing, \ek, $\br(\bsll)$,
$\br(\klpnn)$ and $\br(\kppnn)$ in the model based on supergravity
taking \brbsg\ as one of phenomenological constraints. 
SUSY effects can increase the \bb\ mixing amplitude and \ek\ by up to
20\% in the minimal supergravity model.
If we relax the strict universality of all scalar SUSY breaking masses
at the GUT scale so that we take the soft SUSY breaking scalar masses
for the Higgs fields different from that for squarks and sleptons, the
maximal deviation becomes about 40\%.
The branching ratios for \klpnn\ and \kppnn\ can be suppressed by up to
10\% compared to the SM predictions for the relaxed initial conditions
while the deviation becomes only a few \% in the minimal supergravity
case.

In this paper, we perform quantitative study of FCNC processes in the 
supergravity model taking account of updated constraints on SUSY
parameter space.
Among recent improvements on SUSY searches, the SUSY Higgs boson search
in the LEP~II experiment turns out to give the most significant effect
on constraining the SUSY parameter space.
We show that possible deviation of the \bb\ mixing amplitude from the SM
prediction becomes significantly reduced by the new constraint imposed
by the Higgs boson search, especially for a low \tanb\ region.
Taking into account the recent result of the Higgs boson search, the
maximal deviation of the \bb\ mixing from the SM prediction is 12\% in
the minimal case and 25\% in the nonminimal case for $\tan\beta=3$. If
LEP~II does not find the lightest CP-even Higgs boson below 105 GeV, the 
deviation becomes 11\% and 16\%, respectively.

We also consider the muon anomalous magnetic moment in the supergravity
model.
There have been many works on the muon anomalous magnetic moment in the
MSSM \cite{g-2mssm,Moroi} and in the supergravity model \cite{g-2sugra,CN}.
It is known that this process gives a sizable deviation from the SM for
large \tanb.
At present, a possible new physics contribution to $a_\mu$ is estimated
as $( 75.5 \pm 73.3 ) \times 10^{-10}$ \cite{Marciano}.
For the present Higgs mass bound, we find that the SUSY contribution to
$a_\mu$ can be $(-30$ -- $+80) \times 10^{-10}$ for $\tanb = 10$
and therefore the muon anomalous magnetic moment becomes a useful
constraint on the SUSY parameters when the on-going experiment improves
the sensitivity by a factor of 10--20.

The rest of this paper is organized as follows.
In \refsec{sec:model}, the supergravity model is introduced and the
phenomenological constraints including the Higgs mass bound are
described.
In \refsec{sec:num}, the numerical results on the FCNC processes are
presented.
The numerical results on the muon anomalous magnetic moment are shown in
\refsec{sec:g2mu}.
\refsec{sec:conc} is devoted for the conclusion.

\section{Supergravity model}
\label{sec:model}

In this section we briefly describe the minimal supersymmetric standard
model (MSSM) Lagrangian based on the supergravity model.
Details are given in Refs.~\cite{GNO,bsll,GOS}.

The field contents of MSSM consist of SU(3), SU(2) and U(1) gauge
supermultiplets $(G,\,\wt{G})$, $(W,\,\wt{W})$ and $(B,\,\wt{B})$,
respectively, the Higgs supermultiplets $(h_1,\,\wt{h}_1)$ and
$(h_2,\,\wt{h}_2)$, and chiral supermultiplets corresponding to
quarks/squarks and leptons/sleptons.
The SU(3)$\times$SU(2)$\times$U(1) quantum numbers are given by
\begin{eqnarray}
  Q_i &=& ( 3, \, 2, \, \frac{1}{6} )
~,
~~~
  U_i ~=~ ( \ol{3}, \, 1, \, -\frac{2}{3} )
~,
~~~
  D_i ~=~ ( \ol{3}, \, 1, \, -\frac{2}{3} )
~,
\nonumber\\
  L_i &=& ( 1, \, 2, \, -\frac{1}{2} )
~,
~~~
  E_i ~=~ ( 1, \, 1, \, 1 )
~,
\label{superfield}
\end{eqnarray}
where $i=1,\,2,\,3$ is the generation index.
Requiring the $R$-parity conservation, the MSSM superpotential is given
by
\begin{eqnarray}
  W_{\rm MSSM}
  &=& f_D^{ij} Q_i D_j H_1
    + f_U^{ij} Q_i U_j H_2
    + f_L^{ij} E_i L_j H_1
    + \mu H_1 H_2
~.
\label{superpotential}
\end{eqnarray}
The general form of the soft SUSY breaking terms are given by
\begin{eqnarray}
  -{\cal L}_{\rm soft} &=&
    (m_Q^2)^i_{~j} \wt{q}_i \wt{q}^{\dagger j}
  + (m_D^2)_i^{~j} \wt{d}^{\dagger i} \wt{d}_j
  + (m_U^2)_i^{~j} \wt{u}^{\dagger i} \wt{u}_j
\nonumber\\&&
  + (m_E^2)^i_{~j} \wt{e}_i \wt{e}^{\dagger j} 
  + (m_L^2)_i^{~j} \wt{l}^{\dagger i} \wt{l}_j
\nonumber\\&&
  + \Delta_1^2 h_1^\dagger h_1
  + \Delta_2^2 h_2^\dagger h_2
  - \left( B\mu h_1 h_2 + \hc \right)
\nonumber\\&&
  + \left(
        A_D^{ij} \wt{q}_i \wt{d}_j h_1
      + A_U^{ij} \wt{q}_i \wt{u}_j h_2
      + A_L^{ij} \wt{e}_i \wt{l}_j h_1 + \hc
    \right)
\nonumber\\&&
  + \left(
        \frac{M_1}{2} \wt{B}\wt{B}
      + \frac{M_2}{2} \wt{W}\wt{W}
      + \frac{M_3}{2} \wt{G}\wt{G} + \hc
    \right)
~,
\label{soft}
\end{eqnarray}
where $\wt{q}_i$, $\wt{u}_i$, $\wt{d}_i$,
$\wt{l}_i$, $\wt{e}_i$, $h_1$ and $h_2$  
are scalar components of chiral superfields
$Q_i$, $U_i$, $D_i$,
$L_i$, $E_i$, $H_1$ and $H_2$, respectively, and $\wt{B}$,
$\wt{W}$ and $\wt{G}$ are $U(1)$, $SU(2)$ and $SU(3)$ gauge
fermions, respectively.

In the supergravity model, we assume that the soft SUSY breaking terms
have a simple structure at the Planck or GUT scale.
Here we assume the following initial conditions for renormalization
group equations (RGEs) for the soft SUSY breaking terms at the GUT scale:
\begin{subeqnarray}
  (m_Q^2)^i_{~j} &=&
  (m_E^2)^i_{~j} ~=~ m_0^2\,\delta^i_{~j} ~,
\nonumber\\  
  (m_D^2)_i^{~j} &=&
  (m_U^2)_i^{~j} ~=~
  (m_L^2)_i^{~j} ~=~ m_0^2\,\delta_i^{~j} ~,
\\
  \Delta_1^2 &=& \Delta_2^2 ~=~ \Delta_0^2 ~,
\\
  A_D^{ij} &=& f_{D}^{ij} A_X m_0 ~, ~~ 
  A_L^{ij} ~=~ f_{L}^{ij} A_X m_0 ~, ~~ 
  A_U^{ij} ~=~ f_{U}^{ij} A_X m_0 ~,
\\
  M_1 &=& M_2 ~=~ M_3 ~=~ M_{gX} ~.
\label{softGUT}
\end{subeqnarray}
In the minimal supergravity model the soft breaking parameters $m_0$ and 
$\Delta_0$ are assumed to be equal whereas in the nonminimal case we
treat the two as independent parameters.
We assume that the soft SUSY breaking terms $m_0$, $\Delta_0$, $M_{gX}$, 
$A_X$ and the $\mu$ parameter are all real so that we do not consider the
constraint from electron and neutron electric dipole moments (EDMs).
With the above initial conditions we can solve the one-loop
RGEs for the SUSY breaking parameters and determine these parameters at
the electroweak scale.
We also require that the electroweak symmetry breaking occurs properly
to give the correct $Z^0$ boson mass.
As independent parameters, we take \tanb, $m_0$, $A_X$, $M_{gX}$ and
$\sign(\mu)$ (and $\Delta_0$ for the nonminimal case).
Once we fix a set of these parameters we can calculate the masses and
mixings of all SUSY particles.
We can then evaluate various amplitudes for FCNC processes such as
\brbsg, the \bb\ mixing amplitude, \ek, $\br(\klpnn)$ and
$\br(\kppnn)$.

In order to determine the allowed region of the SUSY parameter space, we 
require the following phenomenological constraints
\renewcommand{\theenumi}{(\arabic{enumi})}
\renewcommand{\labelenumi}{\theenumi}
\begin{enumerate}
\item
  \label{item:bsg}
  \bsg\ constraint from CLEO, \ie,
  $2.0 \times 10^{-4} < \brbsg < 4.5 \times 10^{-4}$ \cite{CLEO}.
\item
  \label{item:lightsp}
  The chargino mass is larger than 97 GeV, and all other charged
  SUSY particle masses should be larger than 90 GeV \cite{LEPII,LP99}.
\item
  All sneutrino masses are larger than 41 GeV \cite{PDG}.
\item
  The gluino and squark mass bounds from TEVATRON experiments
  \cite{TEV}.
\item
  The lightest SUSY particle is neutral.
\item 
  \label{item:ccb}
  The condition for not having a charge or color symmetry breaking
  minimum \cite{ccb}.
\item
  \label{item:higgs}
  SUSY Higgs boson search at LEP~II \cite{LEPhiggs}.
\end{enumerate}
The detail of these conditions \ref{item:bsg}--\ref{item:ccb} is
discussed in Refs.~\cite{GOS,GNO}.
The difference between Refs.~\cite{GOS,GNO} and the present analysis
is that the experimental bounds in \ref{item:bsg} and \ref{item:lightsp} 
are updated and the complete next-to-leading order QCD correction
formula is used for the calculation of the SUSY contributions to the
\bsg\ amplitude~\cite{BMU}.
Previously in Refs.~\cite{GOS,GNO} the condition \ref{item:higgs} did
not give any impact to constrain the SUSY parameter space.
Due to the recent improvement of the Higgs boson search at
LEP~II \cite{LEPhiggs}, a sizable parameter space starts to be excluded.
This will be discussed in the next subsection.

\subsection{Constraint from the SUSY Higgs boson search}
\label{sec:Higgs}

It is well known that there is a strict upper bound on the lightest
CP-even Higgs boson mass in the MSSM \cite{Vloop}. 
Because the Higgs search in the LEP~II experiment has reached to the
sensitivity of 100 GeV for the SM Higgs boson and about 90 GeV for
the case of the lightest CP-even Higgs boson in the MSSM, a meaningful
amount of the SUSY parameter space is already excluded \cite{LEPhiggs}.

The Higgs sector in the MSSM consists of five physical mass eigenstates: 
two CP-even Higgs boson $h$ and $H$, one CP-odd Higgs boson $A$ and a
pair of charged Higgs boson $H^\pm$.
There are two angles to specify the mixing in the Higgs sector, namely, a 
ratio of two vacuum expectation values $\tanb = \vev{h_2^0}/\vev{h_1^0}$
and the mixing angle $\alpha$ for the two CP-even Higgs bosons defined as
\begin{eqnarray}
  \sqrt{2} h_1^0 - v\cos\beta &=& -h \sin\alpha + H \cos\alpha
~,
\nonumber\\
  \sqrt{2} h_2^0 - v\sin\beta &=& \phantom{-}h \cos\alpha + H \sin\alpha
~,
\end{eqnarray}
where we have only kept the CP-even components of two Higgs fields.
In the MSSM, these masses and mixings are determined by two input
parameters, for instance $m_A$ and $m_h$,
as well as stop and sbottom masses and mixing parameters through
one-loop corrections to the Higgs potential formula.
In the supergravity model these masses and mixings at the electroweak
scale are calculated by solving the RGEs from the initial conditions at
the GUT scale.

The constraints on the SUSY parameter space from the LEP experiments are
expressed as an allowed region of two dimensional parameters, for instance
$(m_h,\,m_A)$, with different assumptions of the
remaining parameters (stop mass, \etc).
In order to simplify the treatment in the numerical analysis we use the
following procedure.
We first take the experimental constraint in the space of $m_A$ and
$m_h$ from Ref.~\cite{LEPhiggs,LP99}.
We require that the set of $m_A$ and $m_h$ calculated in the
supergravity model is within the allowed region of this parameter space.
Strictly speaking, the allowed region slightly depends on the stop mass
and mixing parameters through the angles $\alpha$ and $\beta$, but this
dependence is not significant.
In addition, except for a small parameter region in the case of a large
\tanb, the CP-odd scalar mass becomes much larger than $m_Z$ so that the
property of the lightest CP-even Higgs boson becomes very close to that
of the SM Higgs boson.
In such a case the SUSY Higgs boson search essentially gives the same
lower bound of the lightest CP-even Higgs boson mass as that of the SM
Higgs boson, and therefore almost independent of other SUSY parameters.

In the numerical calculation we use the analytic formula for the CP-even 
Higgs mass matrix given in Ref.~\cite{CEQW}.
This formula takes into account two-loop leading-log corrections and
reproduces the next-to-leading-log results within a few GeV for
$m_{\wt{t}}\lsim 1.5$ TeV.
As the present experimental constraint we require that the set of $m_A$
and $m_h$ is within the allowed region shown in \reffig{fig:Higgs} (case 
(I)).
Later we also show how various results depend on a possible future
improvement of the LEP~II Higgs boson search.
We then simply require $m_h\geq 105$ GeV to get a rough idea for the
case that the LEP~II experiment will not discover any light Higgs boson
(case (II)).

\section{Numerical results on FCNC processes}
\label{sec:num}

In this section, the results of numerical calculations for various FCNC
processes in $B$ and $K$ decays are presented.
We consider here
\begin{enumerate}
\item the \bb\ mixing amplitude;
\item \ek;
\item $\br(\klpnn)$;
\item $\br(\kppnn)$.
\end{enumerate}
Detail descriptions on the calculation of these quantities in the
supergravity model are given in Ref.~\cite{GOS,GNO}.
Besides the improvement discussed in \refsec{sec:model}, we have
improved the calculation of the \bb\ and \kk\ mixings by including
the next-to-leading-order QCD correction to the contribution due to the
charged-Higgs--top loop diagram\cite{KKhiggs}.

We first present the magnitude of \dmbd\ in the supergravity
model.
In \reffig{fig:xd-mhl} we show \rxd\ as a function of the lightest Higgs
boson mass for three values of \tanb, \ie\ $\tanb=2,\,3,\,10$, where
$\dmbd^{\rm SM}$ denotes the SM value.
As input parameters we take
$m_t^{\rm pole} = 175$ GeV,
$m_b^{\rm pole} = 4.8$ GeV,
$\alpha_s(m_Z) = 0.119$.
For  the CKM matrix elements, we adopt the 'standard' phase convention
of the Particle Data Group \cite{PDG}, taking $V_{us} = 0.2196$, 
$V_{cb} = 0.0395$, $|V_{ub}/V_{cb}| = 0.08$.
We fix the CP violating phase in the CKM matrix $\delta_{13}=\pi/2$ in
the following analysis.
As we will see later, the ratio \rxd\ is almost independent of
$\delta_{13}$.
The SUSY breaking parameters at the GUT scale are scanned within the
following range:
$0 < m_0 < 1$ TeV,
$0 < \Delta_0 < 1$ TeV,
$0 < M_{gX} < 1$ TeV,
$0 < |A_X| < 5$.
This figure shows the possible values of \rxd\ and $m_h$ for parameter
sets of the minimal supergravity model as well as for those of the
nonminimal case discussed in \refsec{sec:model}.
Campared to the result in \cite{GOS}, we see that  the possible
deviation from the SM is reduced to $+15$\% from $+40$\%  for the
nonminimal case if we require that the lightest 
CP-even Higgs mass should be larger than 95 GeV for $\tanb = 2$. 
If the mass bound is raised to 105 GeV no allowed parameter space
remains for $\tanb = 2$.
For $\tanb = 3$, the lightest Higgs boson mass is shifted to a larger
value so that there is some allowed space for $m_h > 105$ GeV.
For $\tanb = 10$, a separate allowed region appear in the region of
$m_h = 115$--125 GeV where \rxd\ is 1.15--1.30.

It is instructive to see the correlation between \rxd\ and \brbsg.
In \reffig{fig:xd-bsg} we show \rxd\ as a function of \brbsg\ for
$\tanb=2,\,3,\,10$.
In these figures, the constraint from the present SUSY Higgs boson
search is applied as described in \refsec{sec:Higgs}.
For $\tanb=2$, most of the parameter region is already excluded by the
Higgs boson search.
The deviations of both \brbsg\ and \dmbd\ from the SM predictions are 
small in the allowed parameter region.
Compared with the Fig.8(c) in Ref. \cite{GOS},
we can see that the more parameter space is
excluded in the case that \brbsg\ is larger than the SM prediction.
This is because both \brbsg\ and the mass of the lightest CP-even
Higgs boson have a correlation with $\sign(\mu)$. Namely, \brbsg\ 
is enhanced for $\mu<0$.
On the other hannd the lightest Higgs boson becomes lighter for $\mu<0$
since it depends on the $\sign(\mu)$ through the stop left-right mixing
parameter in the radiative correction part of the Higgs mass formula.
For $\tanb = 10$, there are two branches within the allowed region of
\brbsg\ for the nonminimal cese.
The upper dots correspond to the allowed region where $m_h = 115$--125
GeV in \reffig{fig:xd-mhl}(c).
In these points the sign of the \bsg\ amplitude becomes opposite to that 
of the \bsg\ amplitude in the SM because the SUSY contributions are very
large and opposite in sign.
It is known that this parameter space corresponds to the case where the
\bsll\ branching ratio is enhanced by 50--100\%~\cite{bsll}.

In \reffig{fig:xd-tanb}(a) the allowed ranges of \rxd\ are shown for
several values of \tanb\ with the present constraint of the SUSY Higgs
boson search and \reffig{fig:xd-tanb}(b) corresponds to the case where
$m_h > 105$ GeV.
For $\tanb \gsim 10$ there appears a separate region where the sign of
the \bsg\  amplitude is opposite to that of the SM amplitude.
Two cases are distinguished according to the sign of the \bsg\ amplitude
for $\tanb = 10,\,20,\,40$.
We can see that the present constraint allows the deviation of the SM
for the \bb\ mixing up to 12\% for the minimal case and 30\% for the
nonminimal case.
If the Higgs boson bound is raised to 105 GeV, the allowed deviation
from the SM becomes less than 15\% except for the small parameter space
where the sign of the \bsg\ amplitude becomes opposite to the SM case.

Next we show the branching ratio of \klpnn\ normalized to the SM prediction.
We have calculated \rklpnn\ in the minimal and nonminimal cases.
The allowed range of this quantity is shown
for the present Higgs bound (\reffig{fig:klpnn-tanb}(a)) and for $m_h >105$
GeV (\reffig{fig:klpnn-tanb}(b)).
We can see that $\br(\klpnn)$ is suppressed by up to 5\% and 15\% in
the minimal and nonminimal cases, respectively.
When we restrict $m_h > 105$ GeV, the maximal deviation is reduced to 7
\% for the nonminimal case.

We study various correlations among \dmbd, \ek, $\br(\klpnn)$ and
$\br(\kppnn)$.
As noted in Ref.~\cite{GOS}, \rxd\ and \rek\ have a linear relation.
When these quantities are normalized by the corresponding quantities in
the SM, these relations are essentially independent of $\delta_{13}$ in
the CKM matrix because the factors of CKM matrix elements are
approximately equal for \dmbd\ and $\dmbd^{\rm SM}$.
For \ek\ and $\ek^{\rm SM}$, there is a charm loop contribution which
breaks the proportionality, but this contribution turns out to be small.
In \reffig{fig:xd-ek} the correlation between \rxd\ and \rek\ is shown
for $\tanb=3$ and the nonminimal case.
The corresponding figure for the minimal case is obtained by taking a
part of these lines according to the maximal value of \rxd.

The correlation between \rklpnn\ and \rkppnn\ is shown in
\reffig{fig:kppnn-klpnn} for $\tanb=3$ and the nonminimal case.
We can see that the situation is quite similar to \reffig{fig:xd-ek}.
The slight dependence on $\delta_{13}$ is the result of the charm loop
contribution to the $\br(\kppnn)_{\rm SM}$.

We also show the correlation between \rxd\ and \rklpnn\ in
\reffig{fig:klpnn-xd} for $\tanb=3$ and 10 with the present Higgs search 
limit.
We can see that in the parameter space where \dmbd\ is the most
enhanced the suppression of $\br(\klpnn)$ become maximal.
The correlations are useful to distinguish possible new physics effects
from consistency check of the unitarity triangle.
For instance, suppose that the time-dependent CP asymmetry in the
$B\rightarrow J/\psi\,K_S$ decay mode is well established.
Then by combining one more observable quantity, the parameter set
$(\rho,\,\eta)$ of the CKM matrix in the Wolfenstein parametrization is
determined within the SM.
If the SUSY effects are relevant, $|V_{td}|$ determined from
\dmbd\ or \ek\ can be different from that determined from
$\br(\klpnn)$ or $\br(\kppnn)$ because the formers are enhanced and the
latters are suppressed in the supergravity model.
On the other hand, observables such as $|V_{ub}/V_{cb}|$,
$\Delta M_{B_d}/\Delta M_{B_s}$, and CP asymmetries in $B$ decays are
essentially independent of the SUSY loop contributions.

\section{Muon anomalous magnetic moment}
\label{sec:g2mu}

In this section we present the numerical result of SUSY loop effects on
the muon anomalous magnetic moment in the supergravity model.
The present experimental value of $a_\mu = (g-2)_\mu/2$ is \cite{PDG,BNL}
\begin{eqnarray}
  a_\mu^{\rm exp} &=& 11659235.0 (73.0) \times 10^{-10}
~.
\end{eqnarray}
According to Ref. \cite{Marciano} the SM prediction is
\begin{eqnarray}
  a_\mu^{\rm SM} &=& 11659109.6 (6.7) \times 10^{-10}
~,
\end{eqnarray}
where the error in the SM value is dominated by the hadronic
contribution of the vacuum polarization diagram \cite{hadronic,DH}.
Combining the above two values we can derive a possible new physics
contribution to $a_\mu$ as
\begin{eqnarray}
  a_\mu^{\rm exp} - a_\mu^{\rm SM} &=&
  ( 75.5 \pm 73.3 ) \times 10^{-10}
~.
\end{eqnarray}
The current BNL experiment is aiming to improve $a_\mu$ by a factor of
20 and the first result is reported as $a_\mu = 1165925(15) \times
10^{-9}$ \cite{BNL}.
In addition the hadronic contribution to the vacuum polarization may be
better understood if the $e^+ e^-$ total cross section is well measured
experimentally~\cite{DH}.
As we see later, although the present constraint from the muon anomalous 
magnetic moment is not strong enough the situation will soon change
after the improvement of the measurement.

We calculated the SUSY contribution to $a_\mu$ (\amu) from 
the loop diagrams with sneutrino and chargino and with charged slepton
and neutralino. Detailed formula are found, for example, in \cite{Moroi}.
We require the radiative electroweak symmetry breaking condition and
the various phenomenological constraints discussed in \refsec{sec:model}.

We show \amu\ in the minimal supergravity for $\tanb=3$ as a
function of the lighter chargino ($\chi^\pm_1$) mass, the left-handed
smuon ($\wt{\mu}_L$) mass and \brbsg\ in \reffig{fig:g-2mu.03}.
Also the same plots for $\tanb=10$ are shown in \reffig{fig:g-2mu.10}.
In these figures the present bound from the Higgs boson search is
applied.
We can see that \amu\ can be large only when both $\chi^\pm_1$ and 
$\wt{\mu}_L$ are relatively light. For example, $\chi^\pm_1$ and
$\wt{\mu}_L$ are lighter than 150 GeV and 200 GeV, respectively,
in order for $\amu\ > 50 \times 10^{-10}$ for $\tanb = 10$.
As is well known, the magnitude of \amu\ becomes large for
large \tanb.
The enhancement of \amu\ for large \tanb\ comes from the fact that
\amu\ is dominated with the sneutrino-chargino loop diagram, which
contains a contribution proportional to $\mu\tanb$.
As a result, SUSY contributions to both \bsg\ amplitude and
\amu\ are correlated with $\sign(\mu)$.
We can see that \amu\ becomes positive (negative) according
to the suppression (enhancement ) of \brbsg.
This correlation was pointed out in Ref. \cite{CN}.

The predicted range of \amu\ are shown for several values
of \tanb\ with the present constraint on the Higgs search and with
$m_h > 105$ GeV for the minimal and the nonminimal cases in
\reffig{fig:g-2mu-tanb}. As in \reffig{fig:xd-tanb} two cases according
to the sign of the \bsg\  amplitude are shown separately for $\tanb =
10,\, 20,\, 40$. This figure shows that the muon anomalous magnetic
moment is indeed expected to become very powerful to constrain the SUSY
parameter space in near future.
We see that even for $\tanb = 5$ the deviation is quite sizable
considering future improvements on the $a_\mu$ measurement.

\section{Conclusion}
\label{sec:conc}

We updated the numerical analysis of FCNC processes in $B$ and $K$
decays and the muon anomalous magnetic moment in the supergravity model.
Taking account of the recent progress in the Higgs boson search, we show 
that a small \tanb\ region is almost excluded for $\tanb \lsim 2$.
The maximal deviation from the SM value in the \bb\ mixing is 12\% for
the minimal supergravity case and 30\% for the nonminimal case.
If the Higgs mass bound is raised to 105 GeV
the deviations become less than 11\% and 16\%, respectively,
except for a small parameter space where the \bsg\ decay amplitude is
opposite in sign to the SM amplitude.
For $\br(\kpnn)$, we show that the deviation is less than 5\% for the
minimal case and  less than 14\% for the nonminimal case under the
present Higgs mass bound.

We also calculate the SUSY contribution to the muon anomalous magnetic
moment.
\amu\ and \brbsg\ show a strong correlation and \amu\
becomes very large for a large \tanb\ region.
We find that the SUSY contribution \amu\ can be 
$(-30$ -- $+80) \times 10^{-10}$ for $\tanb = 10$ for the minimal
supergravity case. Along with the \brbsg\ constraint, \amu\ will soon
become a very important constraint on the parameter space in the
supergravity model. 

\section*{Acknowledgments}
The authors would like to thank S.~Yamashita for useful discussions on
Higgs boson search experiments at LEP.
The work of Y.O. was supported in part by the Grant-in-Aid of the
Ministry of Education, Science, Sports and Culture, Government of Japan
(No.09640381),Priority area ``Supersymmetry and Unified Theory of
Elementary Particles'' (No. 707), and ``Physics of CP Violation''
(No.09246105). 

\newpage



\clearpage
\section*{Figure Captions}

\newcounter{FIG}
\begin{list}{{\bf FIG. \arabic{FIG}}}{\usecounter{FIG}}
\item 
  Allowed region in $m_A$--$m_h$ space used in our numerical
  calculation.
\label{fig:Higgs}
\item 
  \rxd\ as a function of the lightest Higgs boson mass for (a)
  $\tanb=2$, (b) $\tanb=3$ and (c) $\tanb=10$.
  Each dot represents the value in the full parameter space and
  each square shows the value for the minimal case.
\label{fig:xd-mhl}
\item 
  \rxd\ as a function of \brbsg\ for (a) $\tanb=2$, (b) $\tanb=3$
  and (c) $\tanb=10$.
  Each dot represents the value in the full parameter space and
  each square shows the value for the minimal case.
  The vertical dotted lines show  the upper and lower bounds on
  \brbsg\ given by CLEO.
  The constraint from the present Higgs boson search (case (I)) is
  imposed.
\label{fig:xd-bsg}
\item 
  Allowed ranges of \rxd\ for several values of \tanb\ with 
  (a) the present constraint of the Higgs boson mass (case (I)) and
  (b) $m_h > 105$ GeV (case (II)).
  Two lines are shown according to the sign of the \bsg\ amplitude 
  for $\tanb=10,\,20,\,40$. The left (right) lines correspond the case
  where the sign of the \bsg\ amplitude is same (opposite) as that in
  the SM.
\label{fig:xd-tanb}
\item 
  Allowed ranges of \rklpnn\ for several values of \tanb\ with
  (a) the present constraint of the Higgs boson mass (case (I)) and
  (b) $m_h > 105$ GeV (case (II)).
  The allowed ranges for the parameter regions where the sign of the
  \bsg\ decay amplitude is opposite to that of the SM amplitude are
  separately plotted. The meaning of two lines for  $\tanb=10,\,20,\,40$
  is as the same as in the case of \reffig{fig:xd-tanb}.
\label{fig:klpnn-tanb}
\item 
  Correlation between
  \rxd\ and \rek\
  for
  $\delta_{13}=30^{\circ}$, $90^{\circ}$ and $150^{\circ}$
  in the nonminimal case with $\tanb=3$
  with the present constraint of the Higgs boson mass (case (I)).
\label{fig:xd-ek}
\item 
  Correlation between
  \rklpnn\ and \rkppnn\
  for
  $\delta_{13}=30^{\circ}$, $90^{\circ}$ and $150^{\circ}$
  in the nonminimal case with $\tanb=3$
  with the present constraint of the Higgs boson mass (case (I)).
\label{fig:kppnn-klpnn}
\item 
  Correlation between
  \rklpnn\ and \rxd\ for
  (a) $\tan\beta = 3$ and
  (b) $\tan\beta = 10$
  with the present constraint of the Higgs boson mass (case (I)).
\label{fig:klpnn-xd}
\item 
  \amu\ in the minimal supergravity case for $\tanb=3$
  (a) as a function of the lighter chargino mass,
  (b) as a function of the left-handed scalar muon mass,
  and
  (c) as a function of \brbsg.
  The constraint from the present Higgs boson search (case (I)) is
  imposed.
\label{fig:g-2mu.03}
\item 
  Same as \reffig{fig:g-2mu.03} for $\tanb=10$.
\label{fig:g-2mu.10}
\item 
  Allowed ranges of \amu\ for several values of \tanb\ with 
  (a) the present constraint of the Higgs boson mass (case (I)) and
  (b) $m_h > 105$ GeV (case (II)).
  The allowed ranges for the parameter regions where the sign of the
  \bsg\ decay amplitude is opposite to that of the SM amplitude are
  separately plotted.
\label{fig:g-2mu-tanb}
\end{list}

\clearpage
\section*{Figures}
\pagestyle{empty}

\def\EPSDIR{}
\def\EPSSCALE{0.9}

~
\vfill
\begin{center}
\makebox[0cm]{
\def\epsfsize#1#2{\EPSSCALE#1}
\epsfbox{\EPSDIR 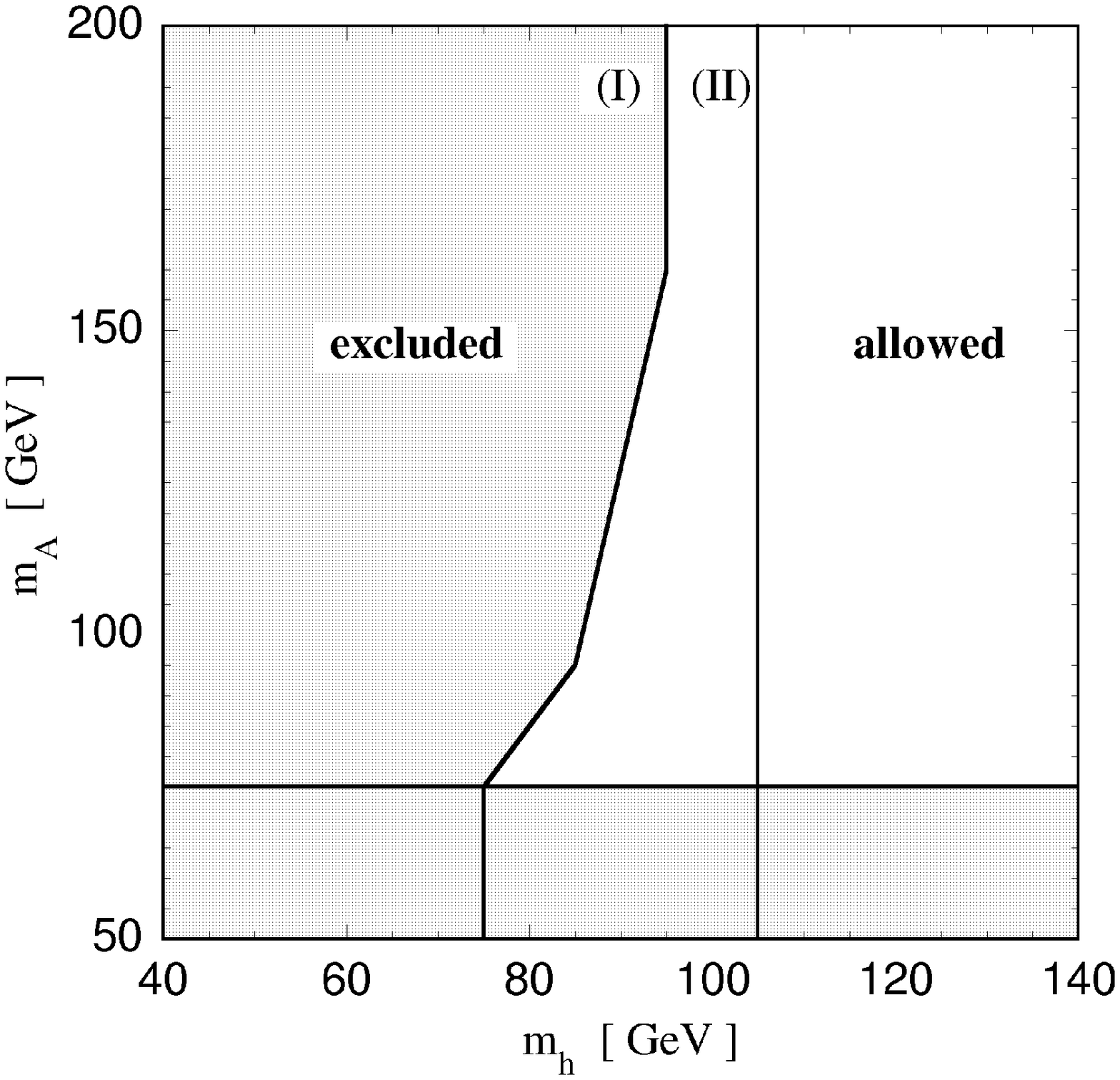}
}
\vfill
{\Large\bf Fig.~\ref{fig:Higgs}}
\end{center}
\clearpage

~
\vfill
\begin{center}
\makebox[0cm]{
\def\epsfsize#1#2{\EPSSCALE#1}
\epsfbox{\EPSDIR 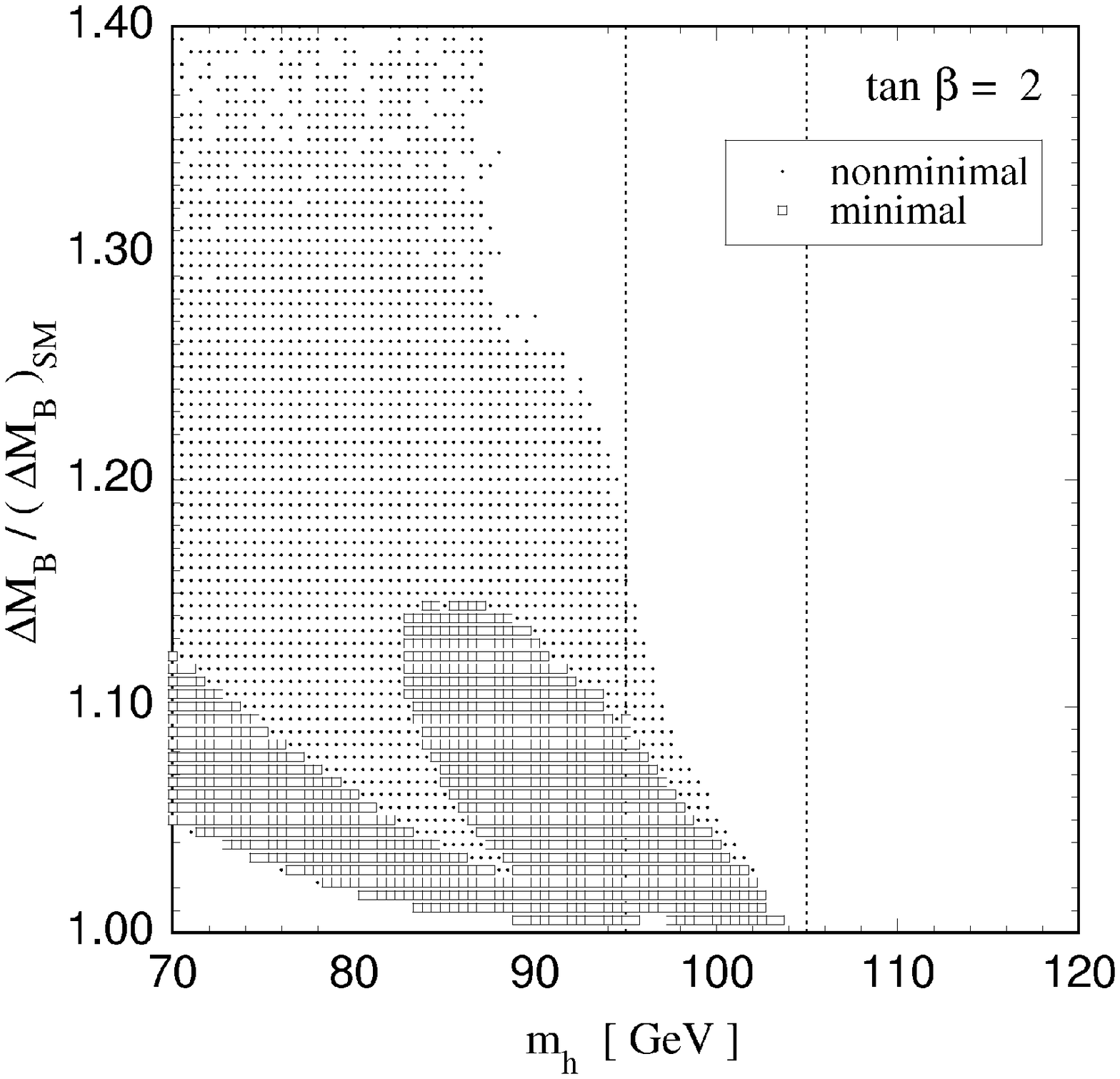}
}
\vfill
{\Large\bf Fig.~\ref{fig:xd-mhl}(a)}
\end{center}
\clearpage

~
\vfill
\begin{center}
\makebox[0cm]{
\def\epsfsize#1#2{\EPSSCALE#1}
\epsfbox{\EPSDIR 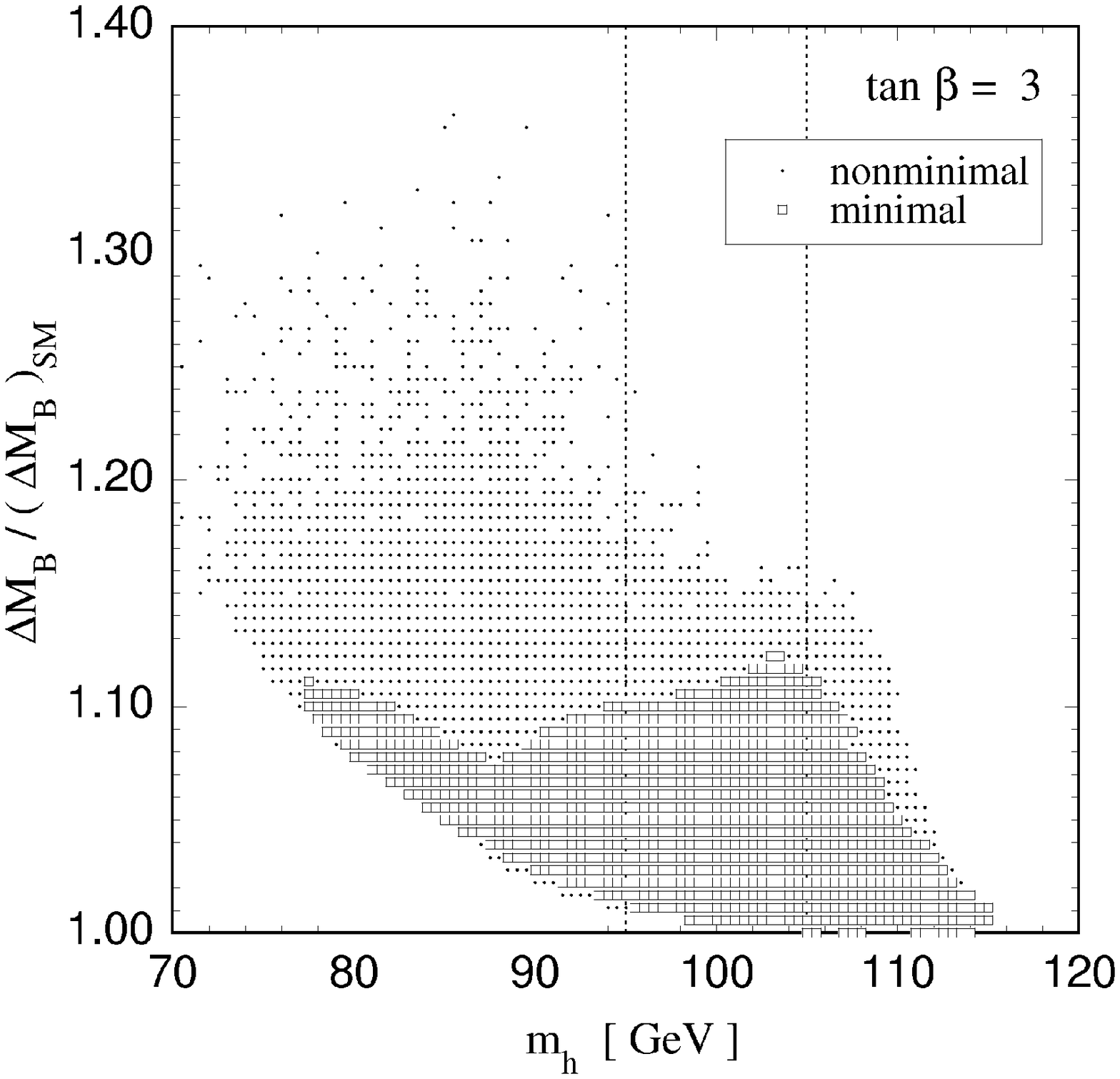}
}
\vfill
{\Large\bf Fig.~\ref{fig:xd-mhl}(b)}
\end{center}
\clearpage

~
\vfill
\begin{center}
\makebox[0cm]{
\def\epsfsize#1#2{\EPSSCALE#1}
\epsfbox{\EPSDIR 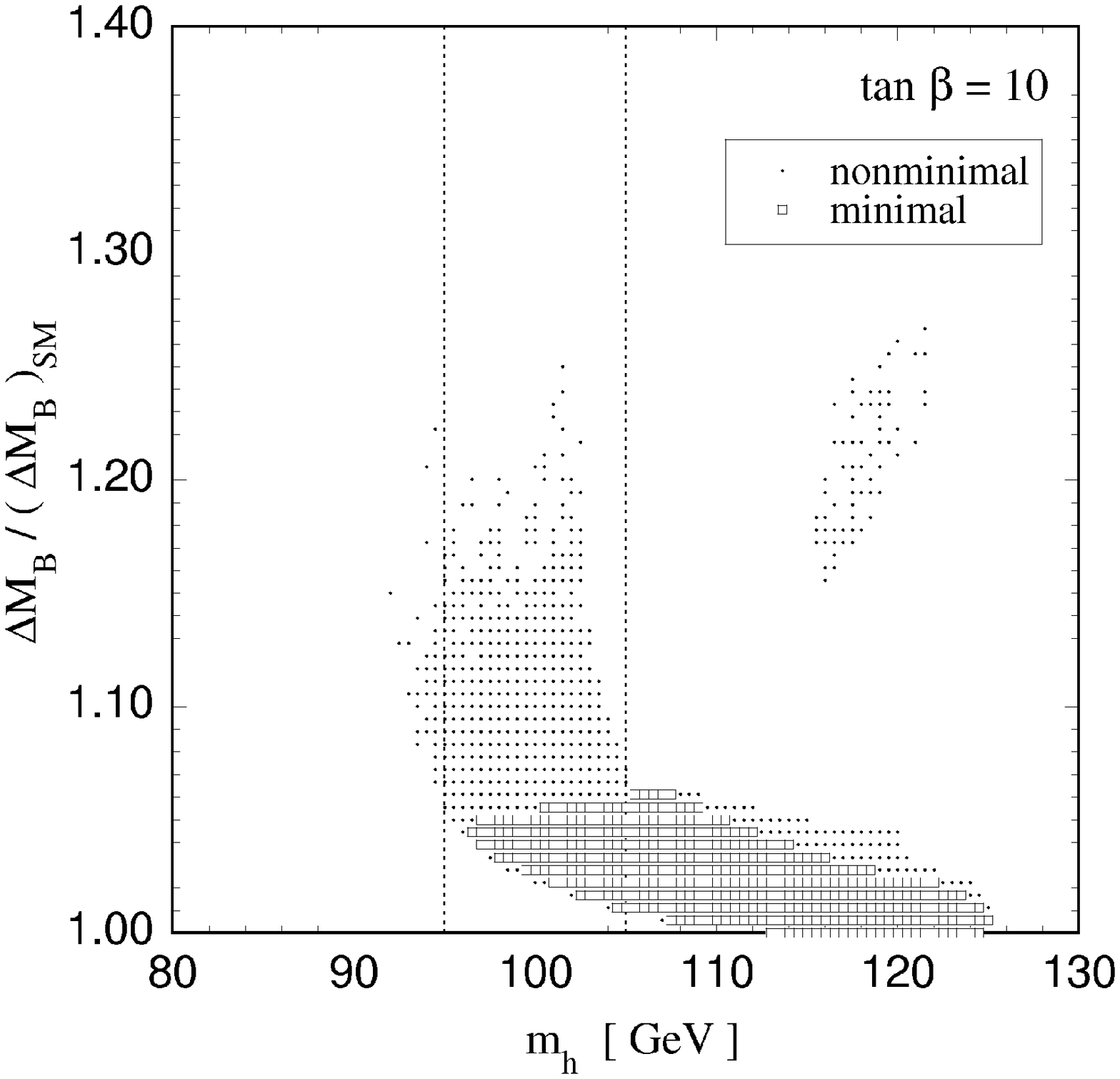}
}
\vfill
{\Large\bf Fig.~\ref{fig:xd-mhl}(c)}
\end{center}
\clearpage

~
\vfill
\begin{center}
\makebox[0cm]{
\def\epsfsize#1#2{\EPSSCALE#1}
\epsfbox{\EPSDIR 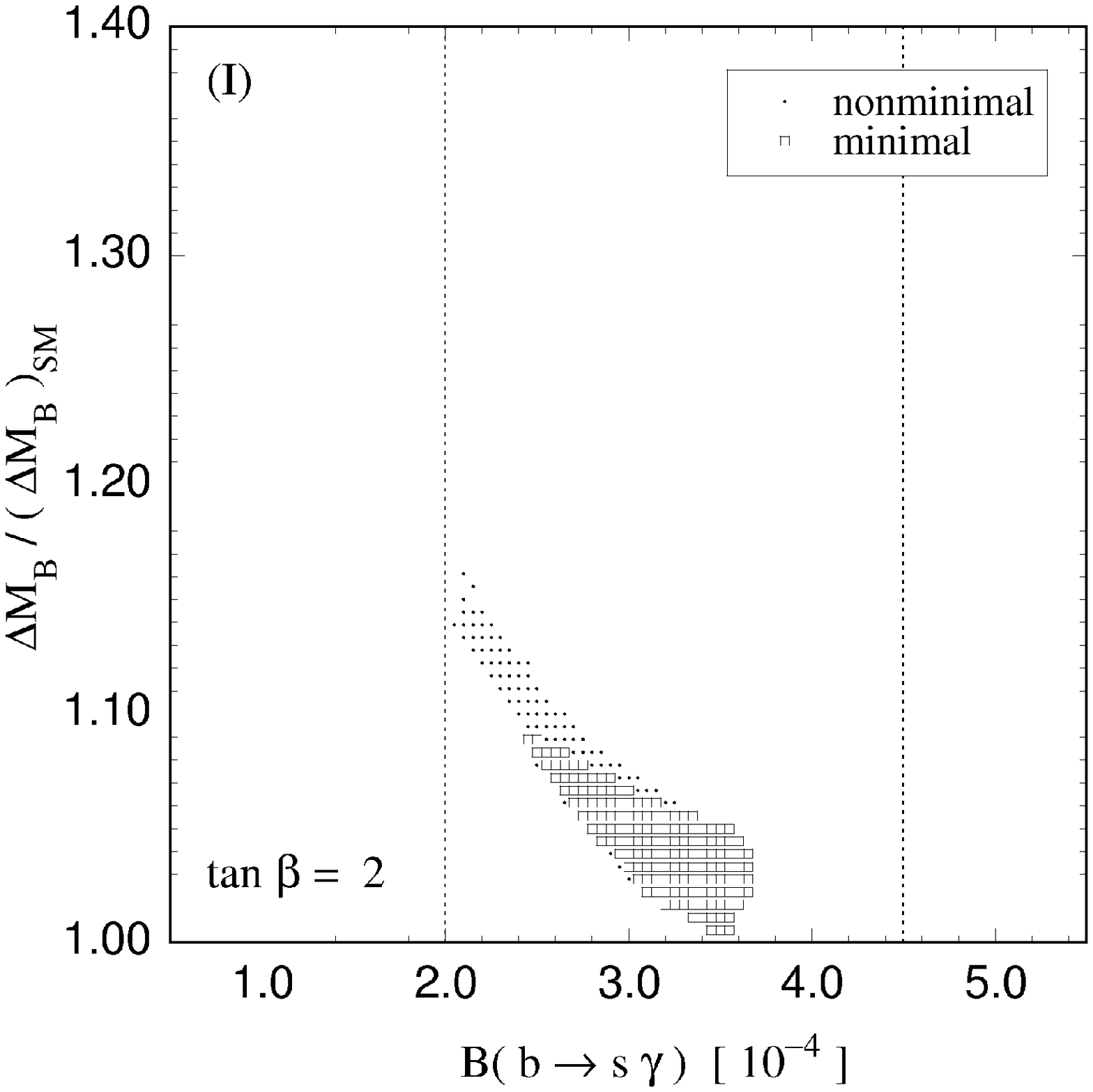}
}
\vfill
{\Large\bf Fig.~\ref{fig:xd-bsg}(a)}
\end{center}
\clearpage

~
\vfill
\begin{center}
\makebox[0cm]{
\def\epsfsize#1#2{\EPSSCALE#1}
\epsfbox{\EPSDIR 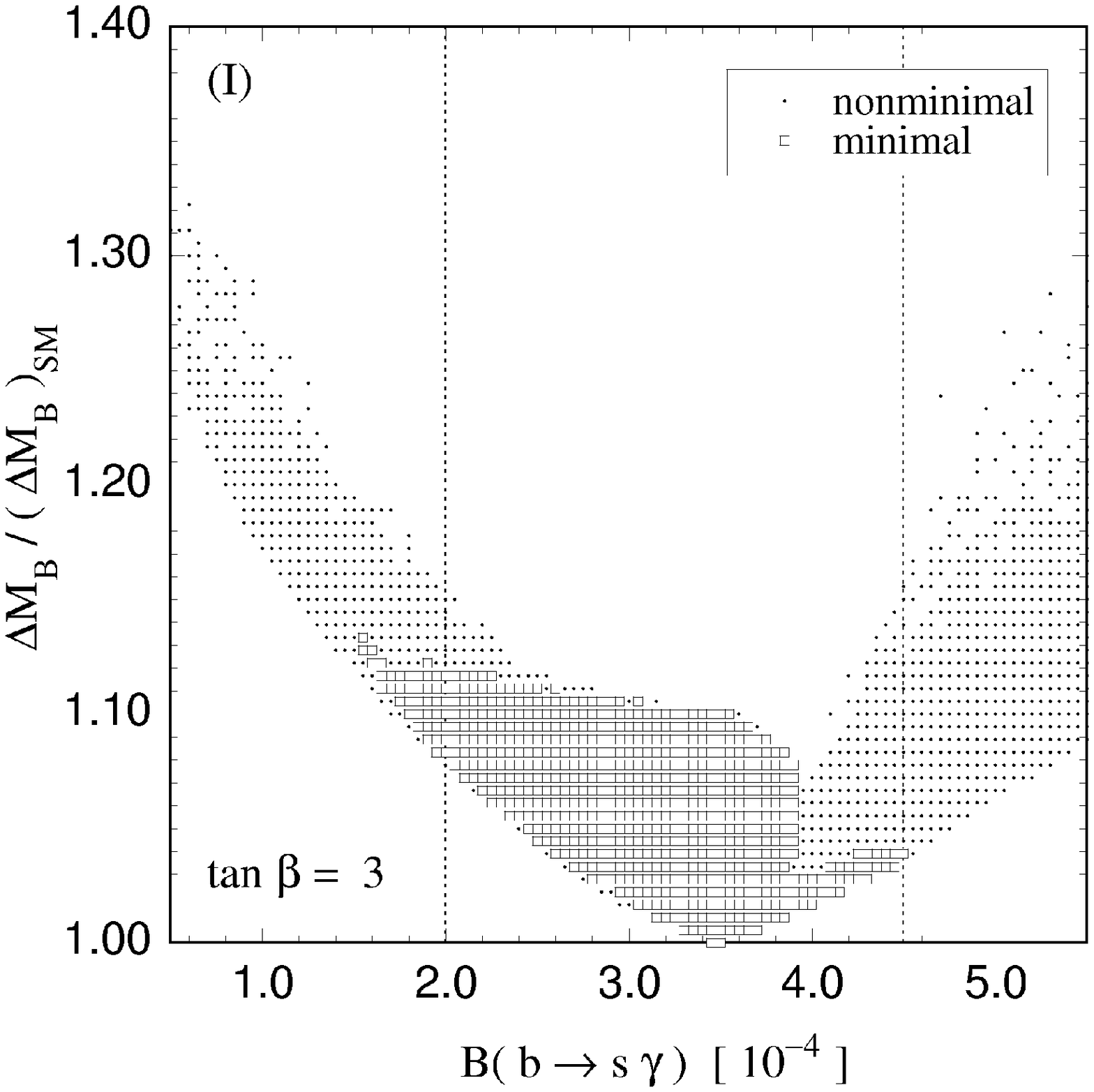}
}
\vfill
{\Large\bf Fig.~\ref{fig:xd-bsg}(b)}
\end{center}
\clearpage

~
\vfill
\begin{center}
\makebox[0cm]{
\def\epsfsize#1#2{\EPSSCALE#1}
\epsfbox{\EPSDIR 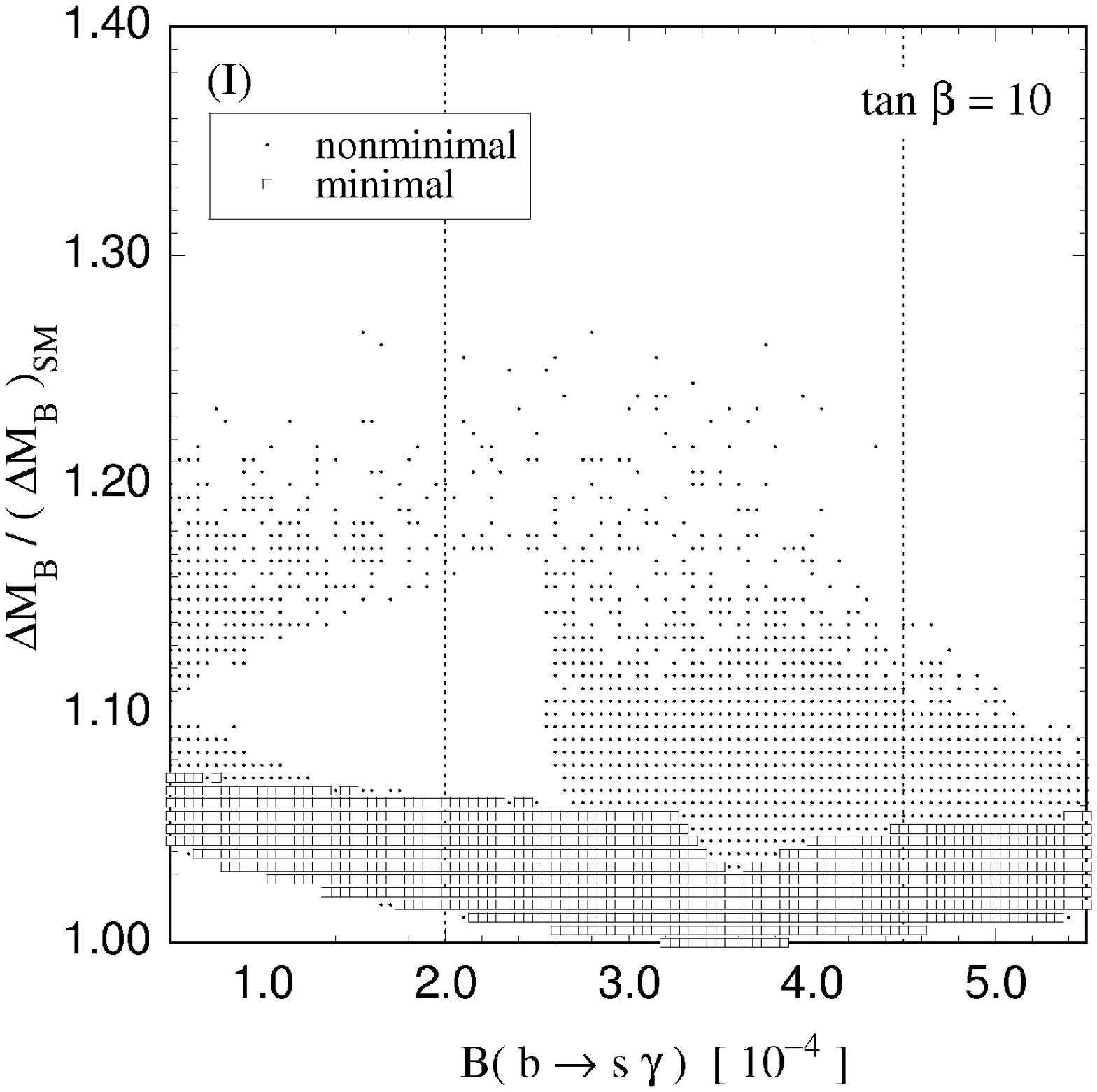}
}
\vfill
{\Large\bf Fig.~\ref{fig:xd-bsg}(c)}
\end{center}
\clearpage

~
\vfill
\begin{center}
\makebox[0cm]{
\def\epsfsize#1#2{\EPSSCALE#1}
\epsfbox{\EPSDIR 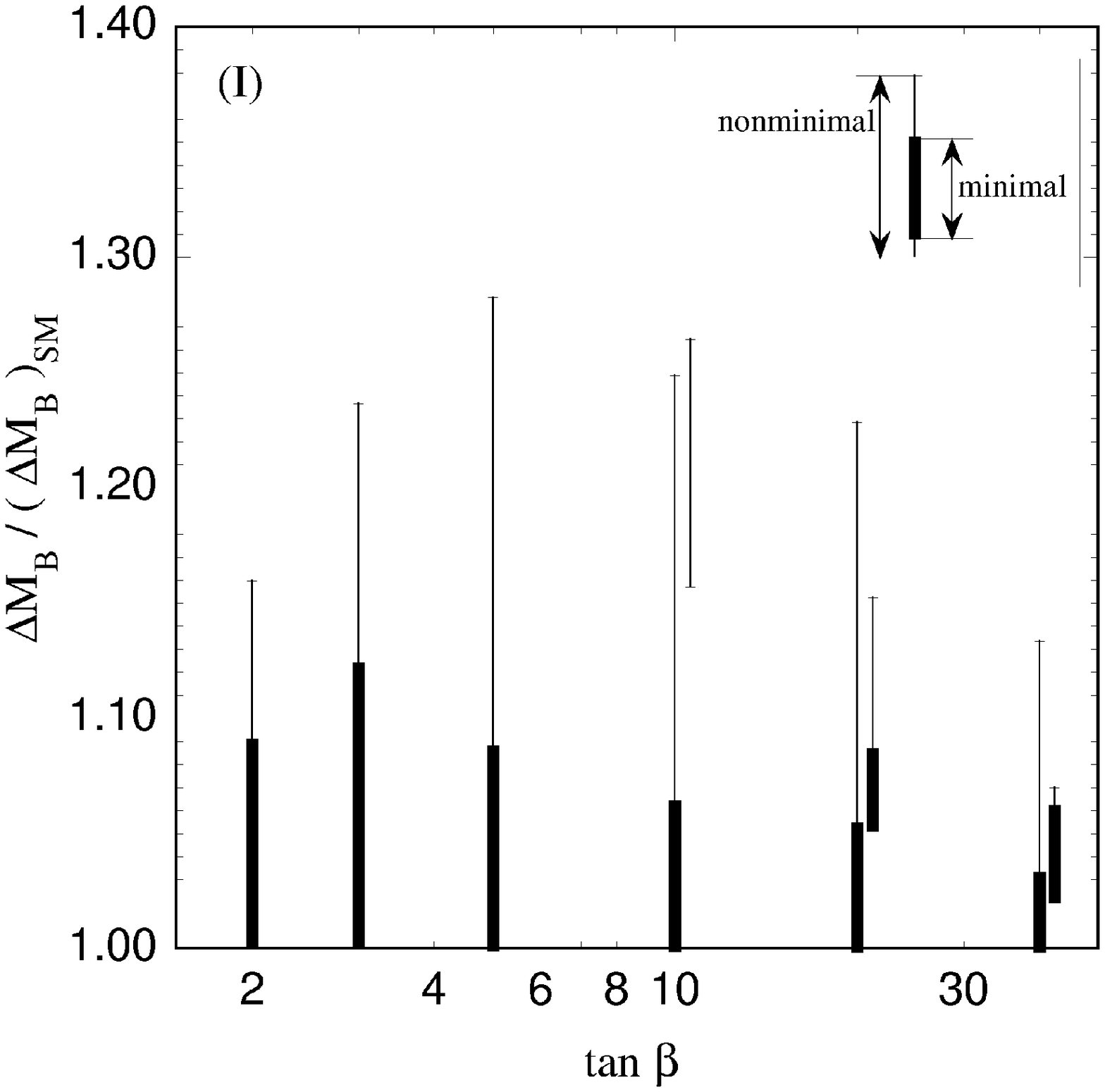}
}
\vfill
{\Large\bf Fig.~\ref{fig:xd-tanb}(a)}
\end{center}
\clearpage

~
\vfill
\begin{center}
\makebox[0cm]{
\def\epsfsize#1#2{\EPSSCALE#1}
\epsfbox{\EPSDIR 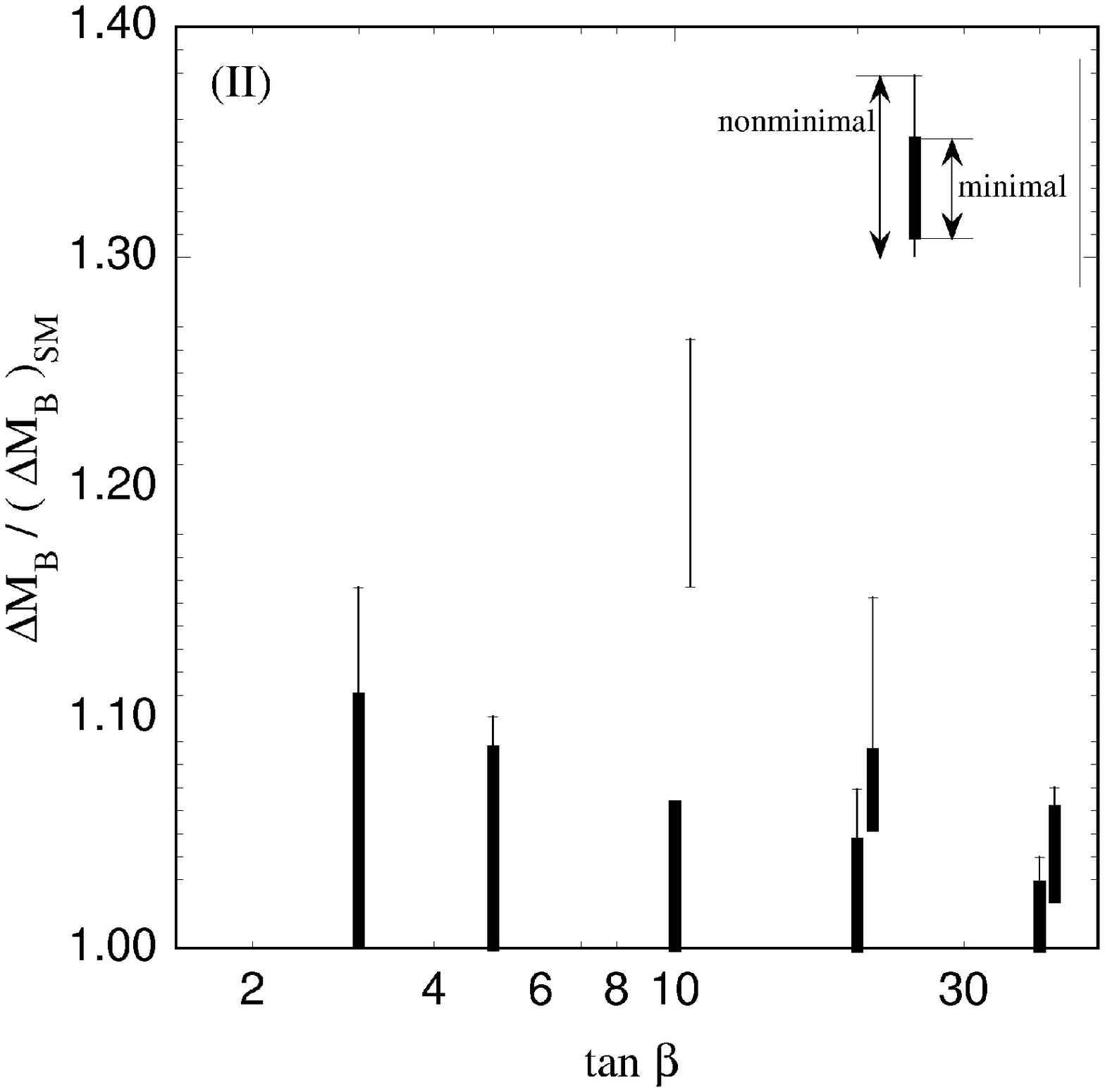}
}
\vfill
{\Large\bf Fig.~\ref{fig:xd-tanb}(b)}
\end{center}
\clearpage

~
\vfill
\begin{center}
\makebox[0cm]{
\def\epsfsize#1#2{\EPSSCALE#1}
\epsfbox{\EPSDIR 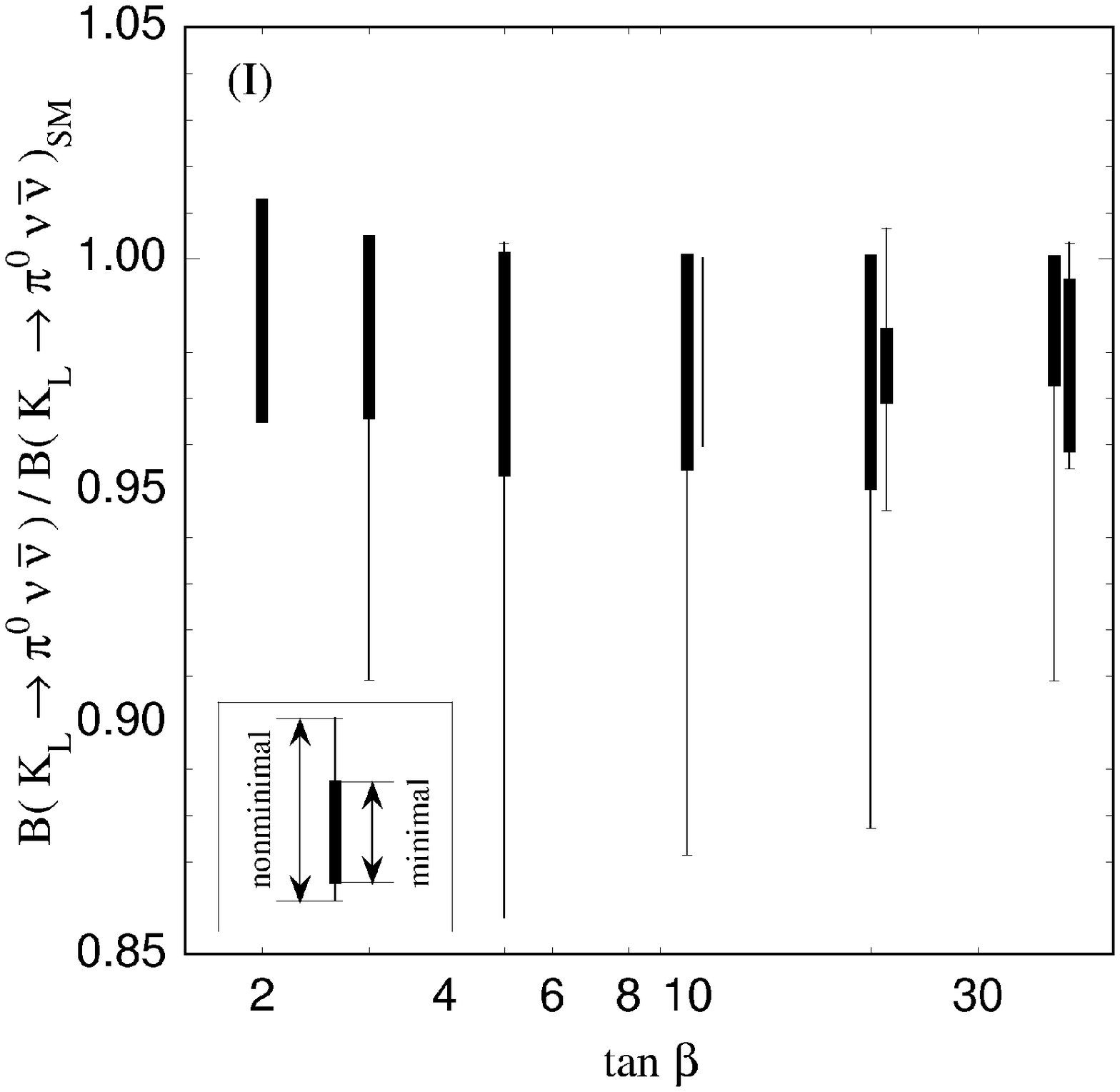}
}
\vfill
{\Large\bf Fig.~\ref{fig:klpnn-tanb}(a)}
\end{center}
\clearpage

~
\vfill
\begin{center}
\makebox[0cm]{
\def\epsfsize#1#2{\EPSSCALE#1}
\epsfbox{\EPSDIR 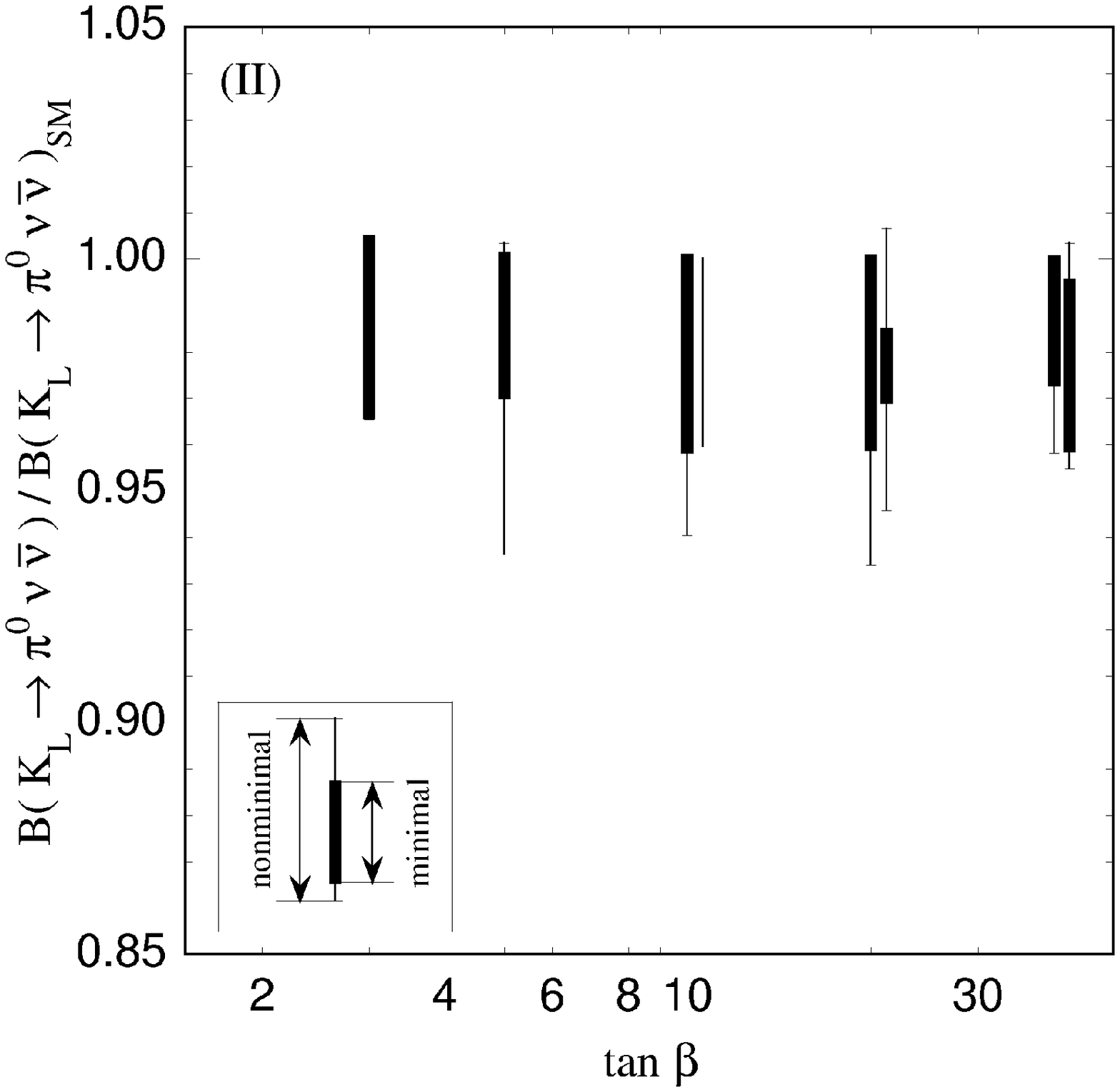}
}
\vfill
{\Large\bf Fig.~\ref{fig:klpnn-tanb}(b)}
\end{center}
\clearpage

~
\vfill
\begin{center}
\makebox[0cm]{
\def\epsfsize#1#2{\EPSSCALE#1}
\epsfbox{\EPSDIR 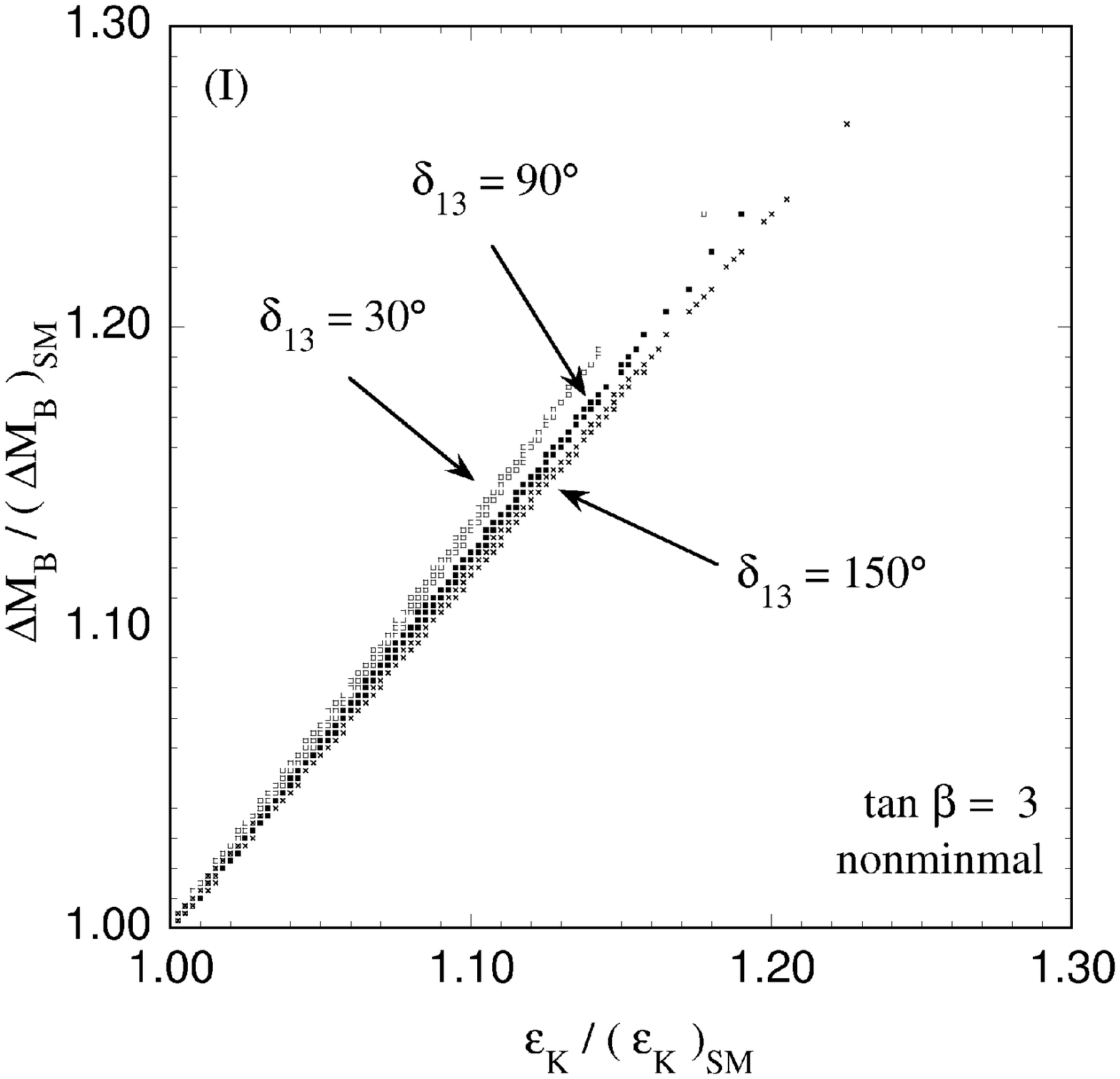}
}
\vfill
{\Large\bf Fig.~\ref{fig:xd-ek}}
\end{center}
\clearpage

~
\vfill
\begin{center}
\makebox[0cm]{
\def\epsfsize#1#2{\EPSSCALE#1}
\epsfbox{\EPSDIR 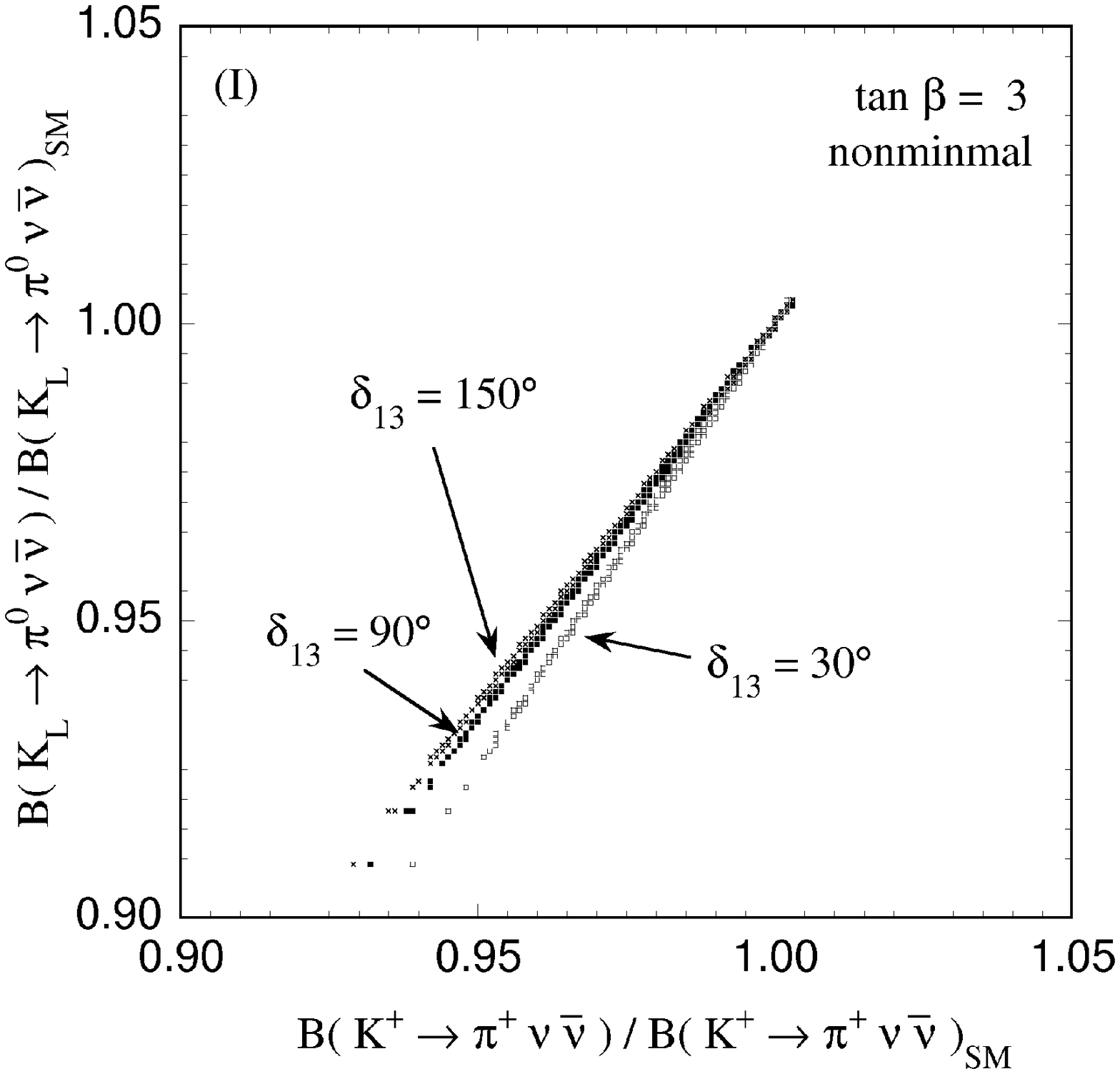}
}
\vfill
{\Large\bf Fig.~\ref{fig:kppnn-klpnn}}
\end{center}
\clearpage

~
\vfill
\begin{center}
\makebox[0cm]{
\def\epsfsize#1#2{\EPSSCALE#1}
\epsfbox{\EPSDIR 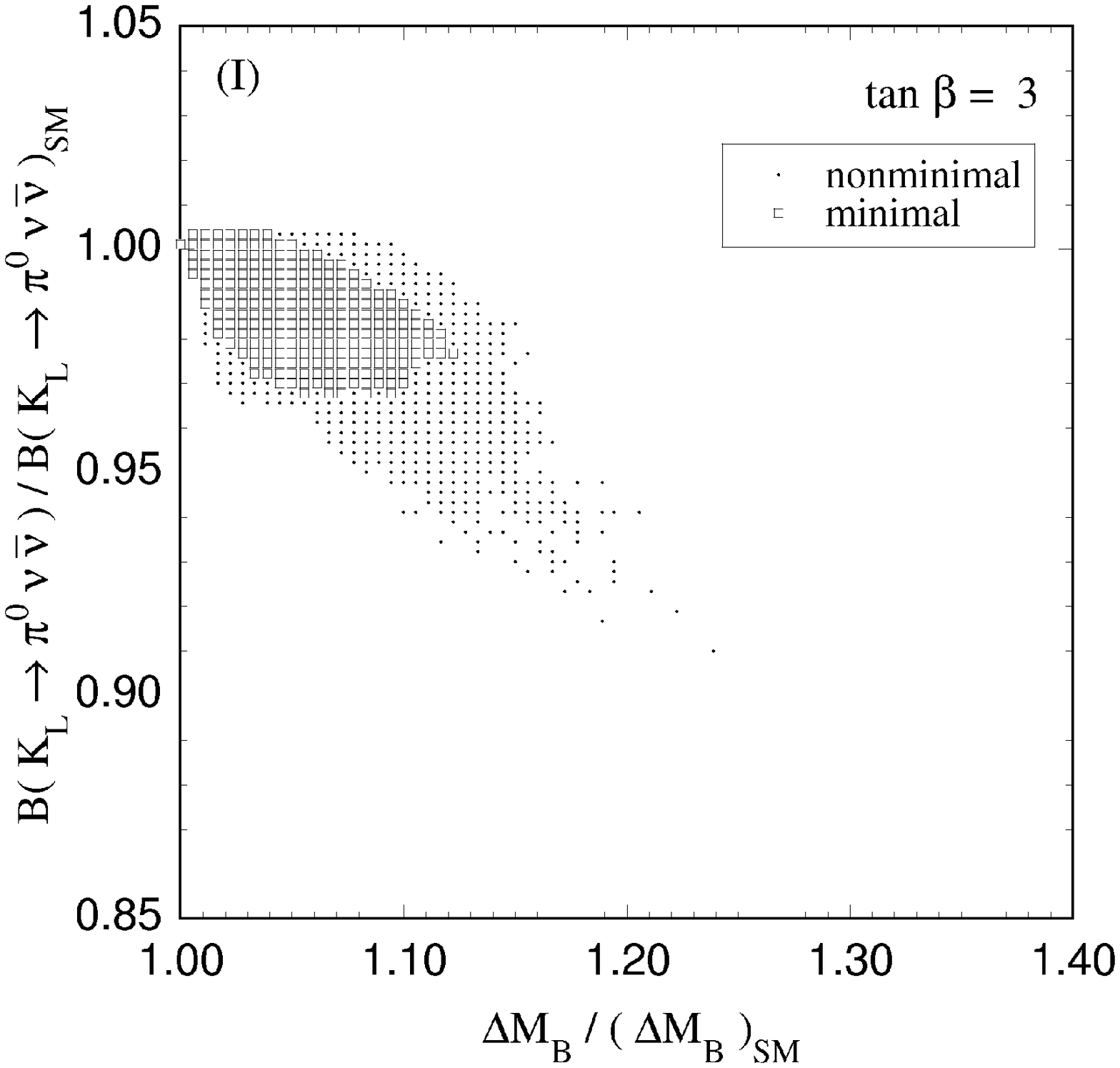}
}
\vfill
{\Large\bf Fig.~\ref{fig:klpnn-xd}(a)}
\end{center}
\clearpage

~
\vfill
\begin{center}
\makebox[0cm]{
\def\epsfsize#1#2{\EPSSCALE#1}
\epsfbox{\EPSDIR 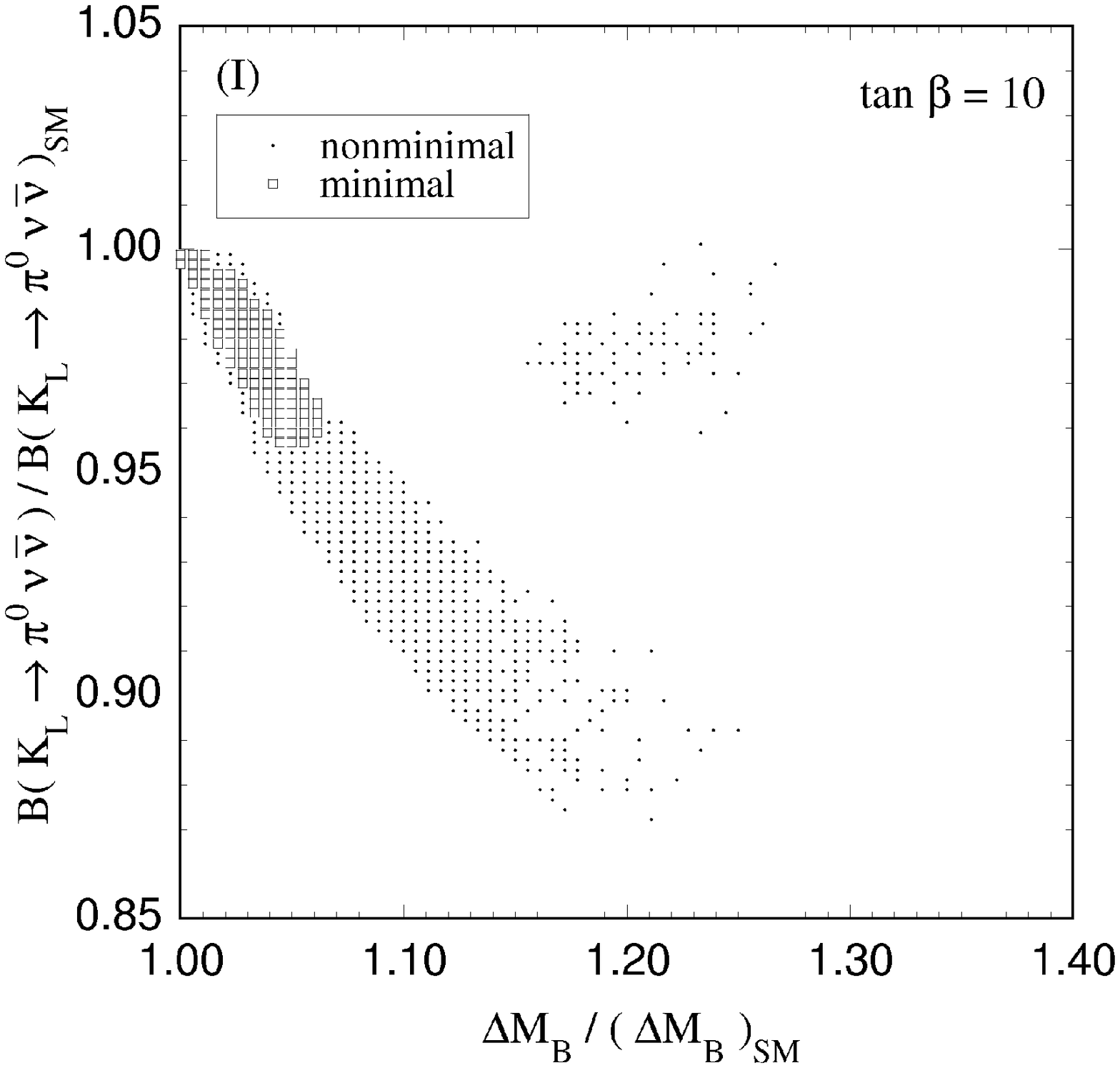}
}
\vfill
{\Large\bf Fig.~\ref{fig:klpnn-xd}(b)}
\end{center}
\clearpage

~
\vfill
\begin{center}
\makebox[0cm]{
\def\epsfsize#1#2{\EPSSCALE#1}
\epsfbox{\EPSDIR 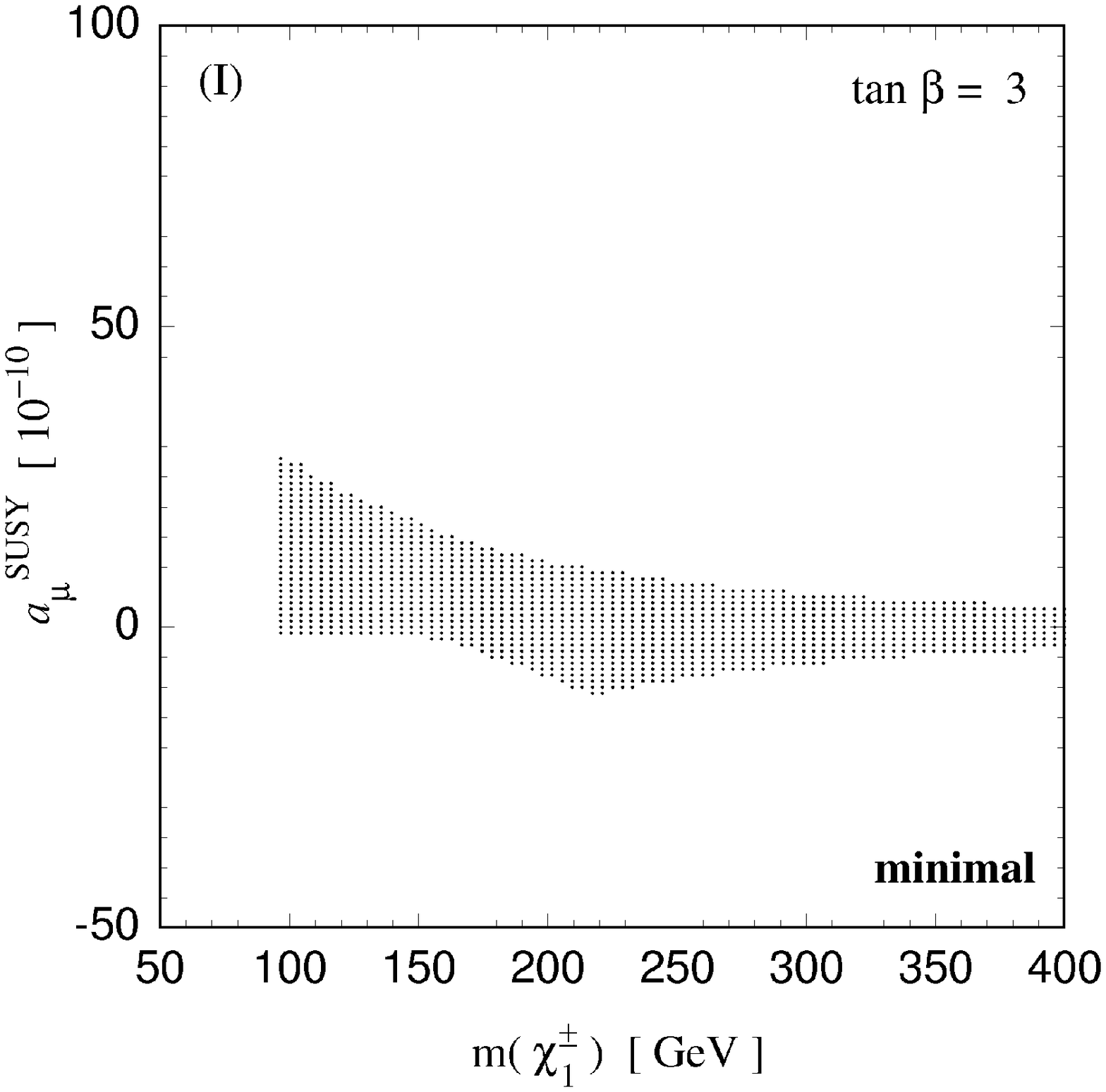}
}
\vfill
{\Large\bf Fig.~\ref{fig:g-2mu.03}(a)}
\end{center}
\clearpage

~
\vfill
\begin{center}
\makebox[0cm]{
\def\epsfsize#1#2{\EPSSCALE#1}
\epsfbox{\EPSDIR 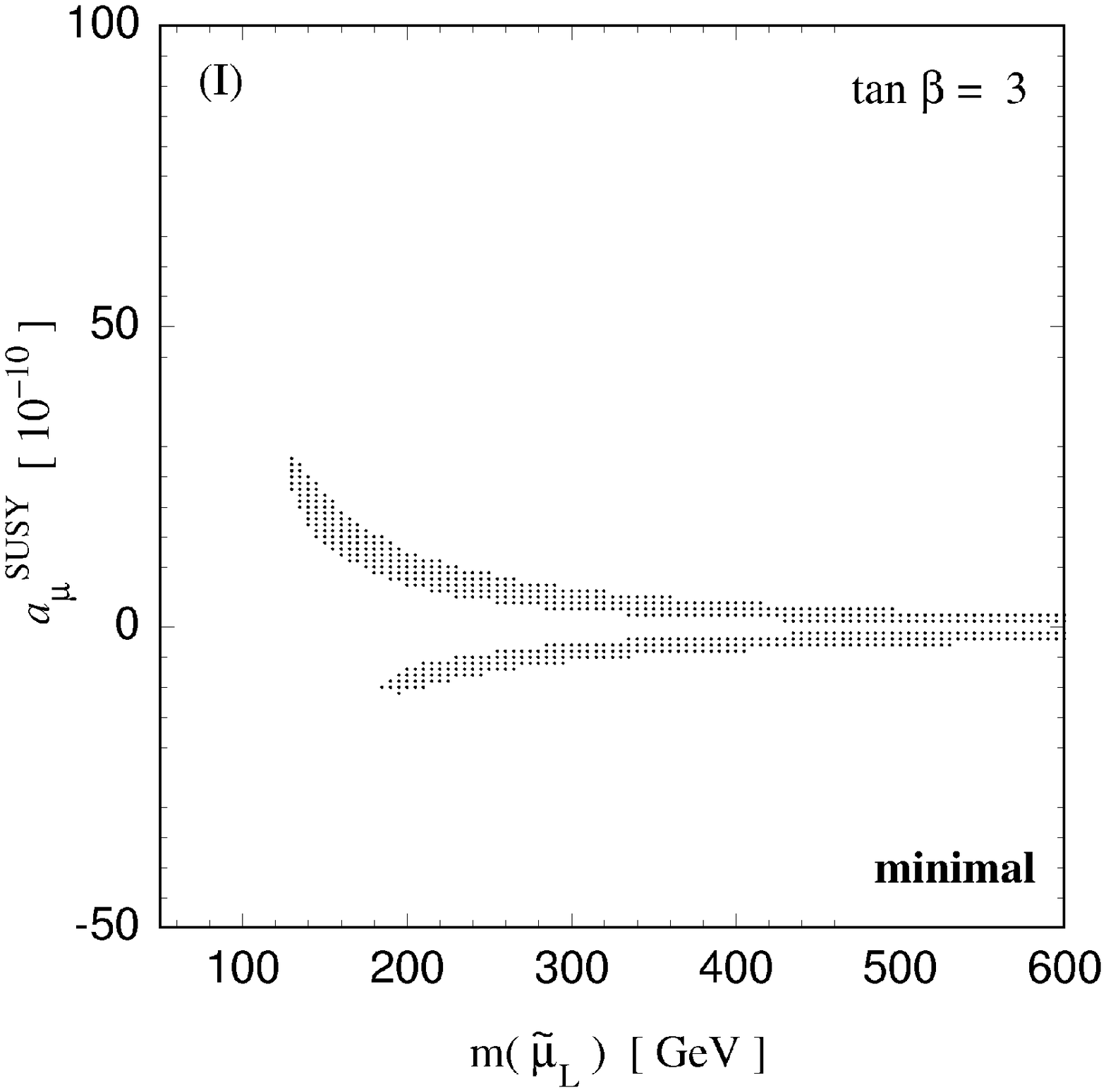}
}
\vfill
{\Large\bf Fig.~\ref{fig:g-2mu.03}(b)}
\end{center}
\clearpage

~
\vfill
\begin{center}
\makebox[0cm]{
\def\epsfsize#1#2{\EPSSCALE#1}
\epsfbox{\EPSDIR 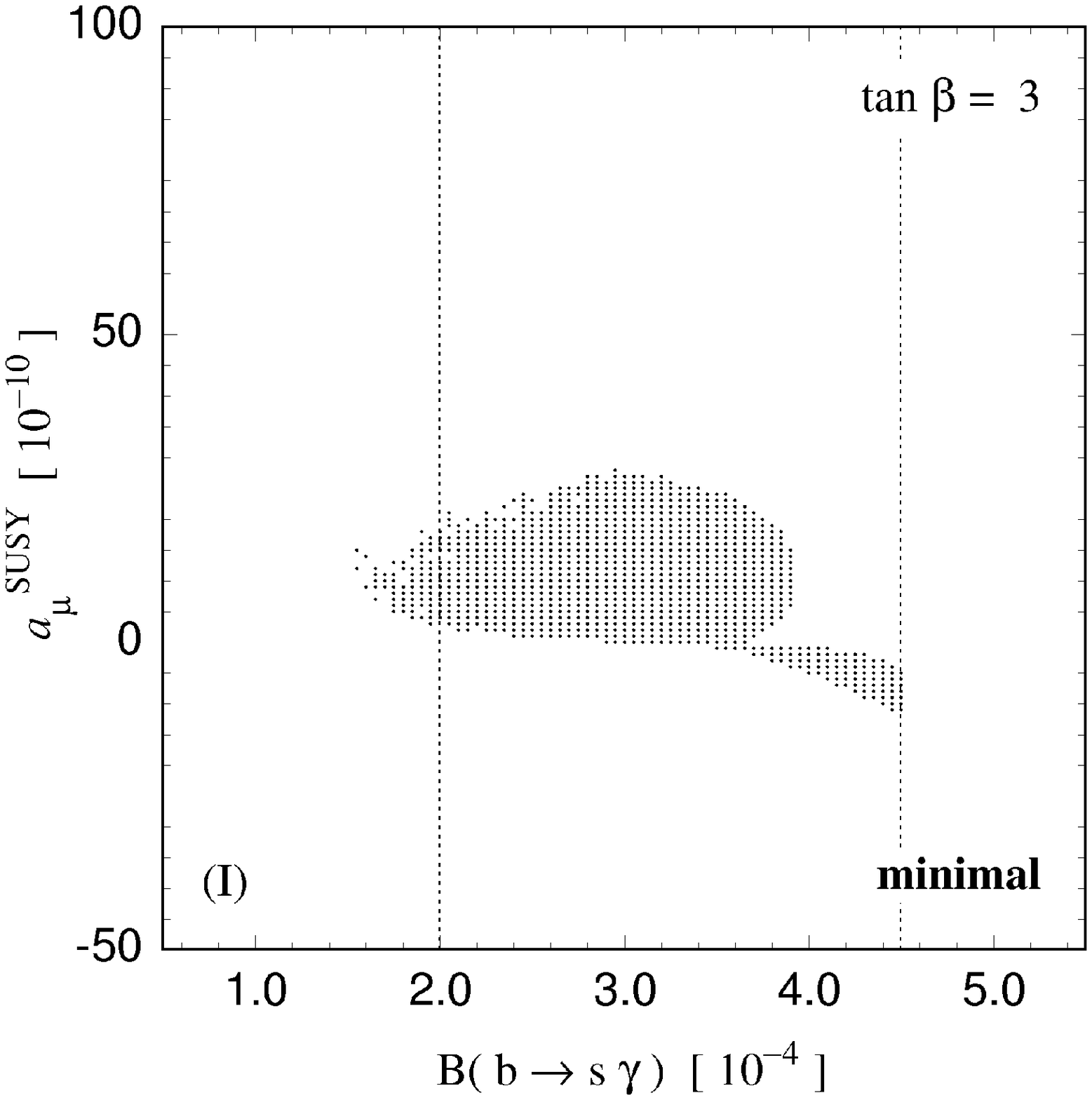}
}
\vfill
{\Large\bf Fig.~\ref{fig:g-2mu.03}(c)}
\end{center}
\clearpage

~
\vfill
\begin{center}
\makebox[0cm]{
\def\epsfsize#1#2{\EPSSCALE#1}
\epsfbox{\EPSDIR 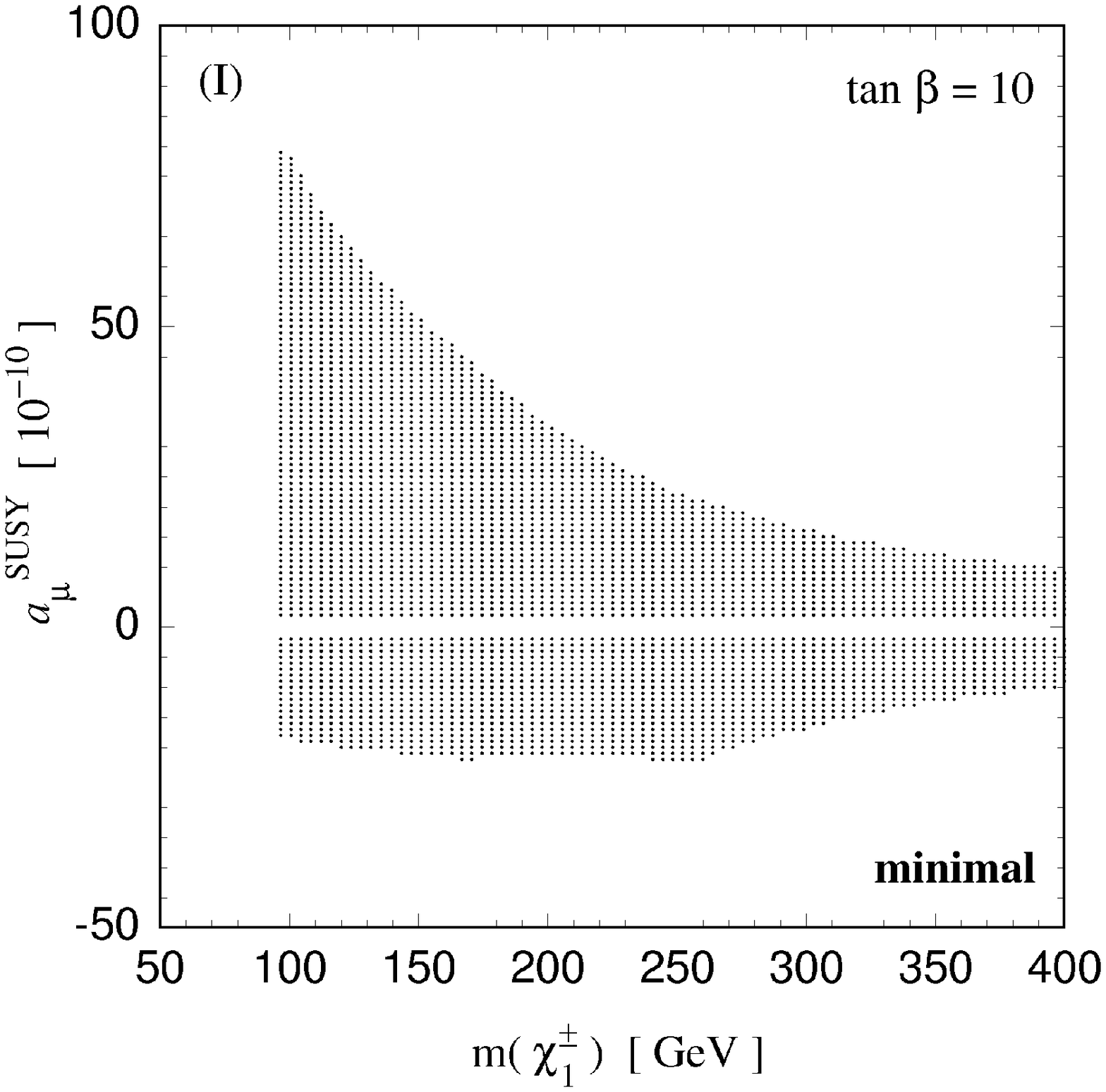}
}
\vfill
{\Large\bf Fig.~\ref{fig:g-2mu.10}(a)}
\end{center}
\clearpage

~
\vfill
\begin{center}
\makebox[0cm]{
\def\epsfsize#1#2{\EPSSCALE#1}
\epsfbox{\EPSDIR 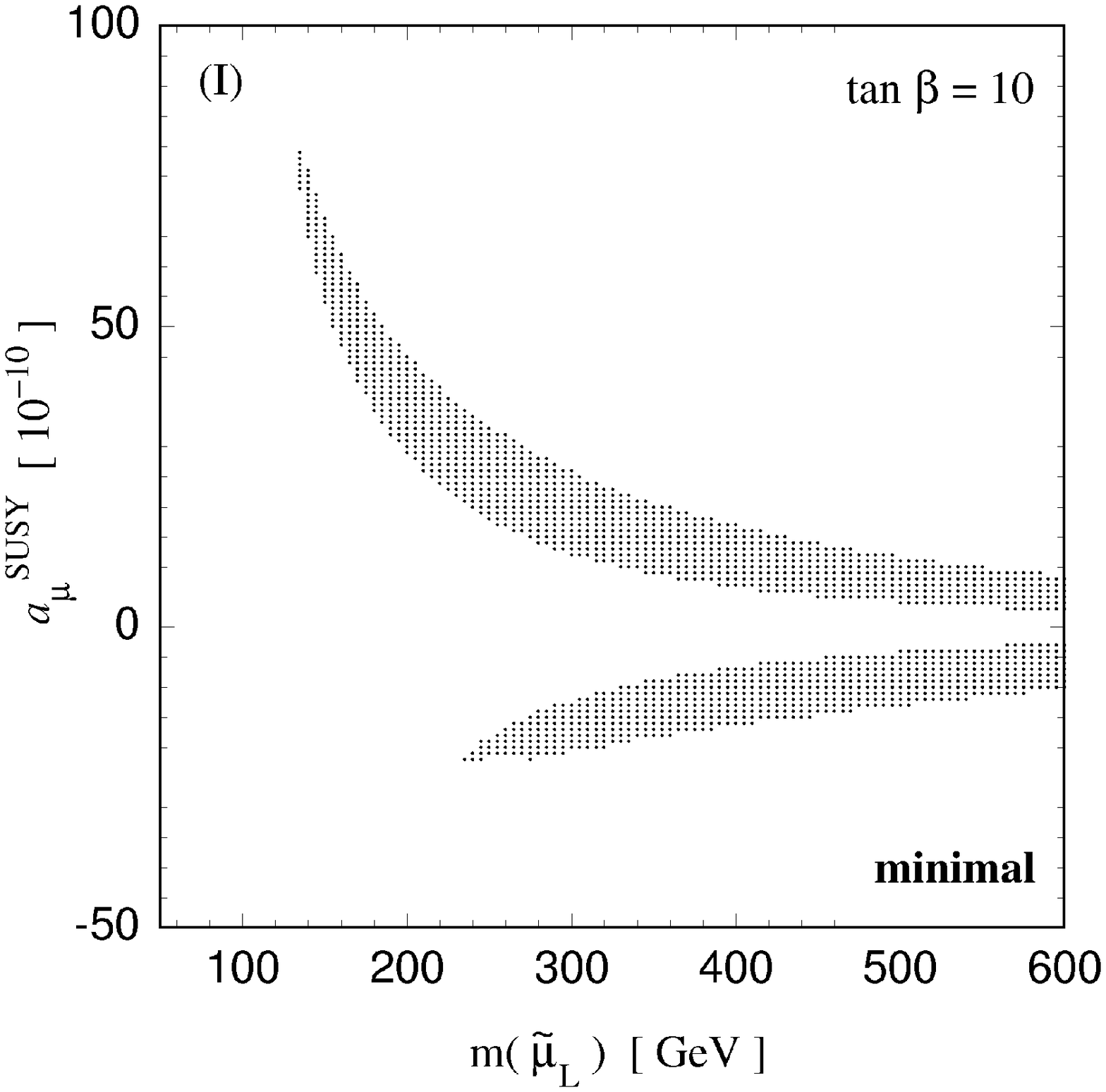}
}
\vfill
{\Large\bf Fig.~\ref{fig:g-2mu.10}(b)}
\end{center}
\clearpage

~
\vfill
\begin{center}
\makebox[0cm]{
\def\epsfsize#1#2{\EPSSCALE#1}
\epsfbox{\EPSDIR 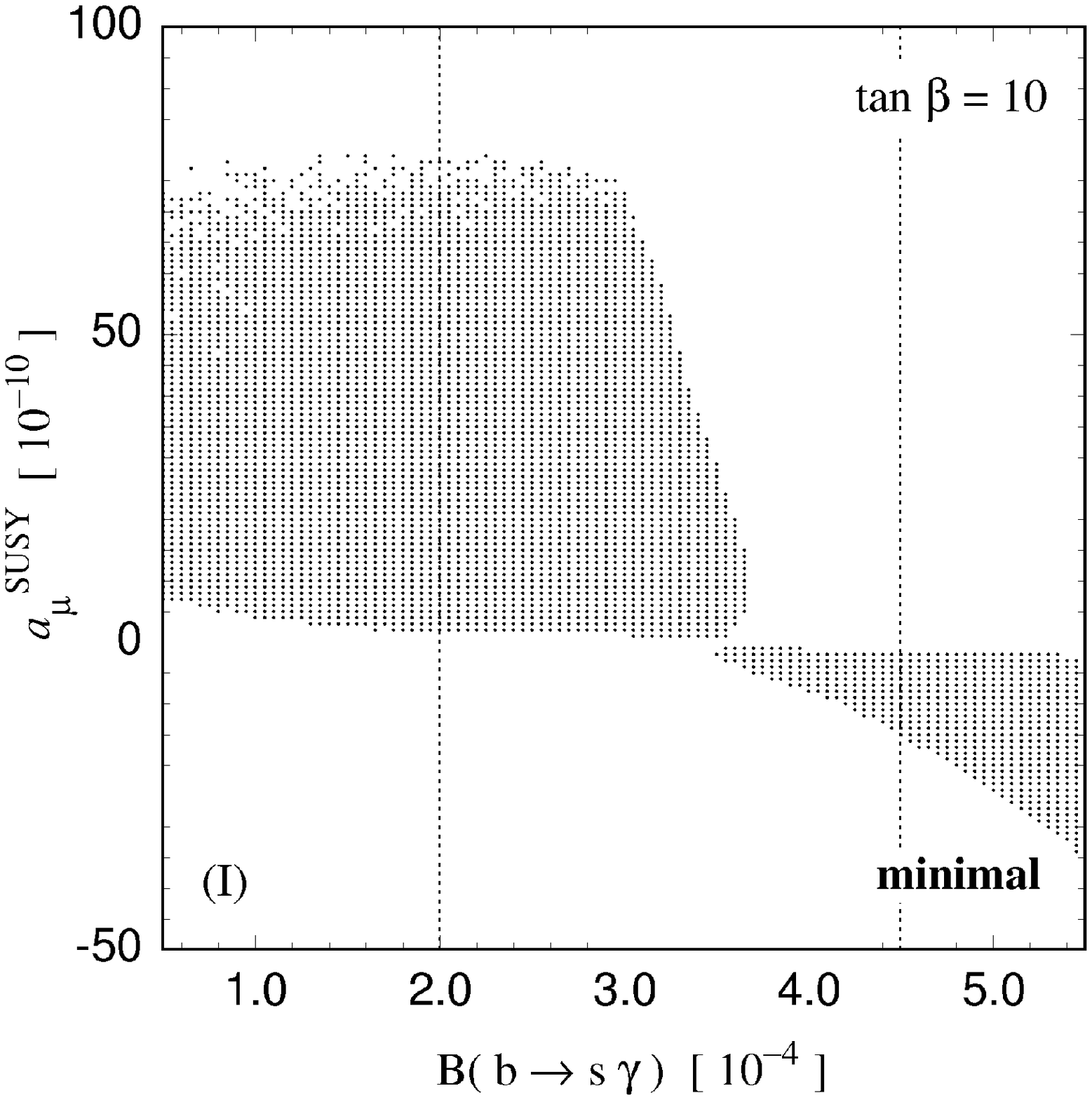}
}
\vfill
{\Large\bf Fig.~\ref{fig:g-2mu.10}(c)}
\end{center}
\clearpage

~
\vfill
\begin{center}
\makebox[0cm]{
\def\epsfsize#1#2{\EPSSCALE#1}
\epsfbox{\EPSDIR 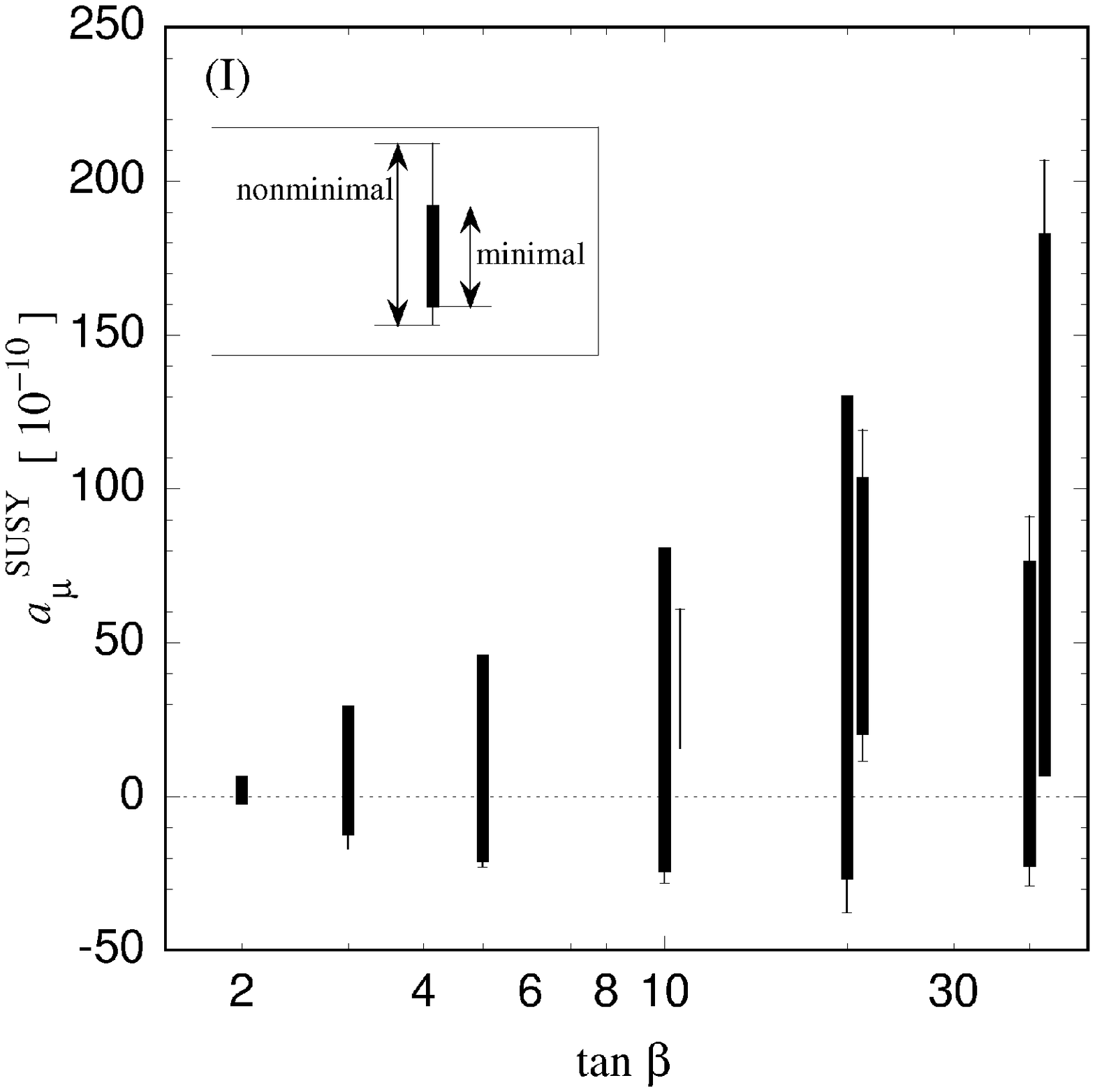}
}
\vfill
{\Large\bf Fig.~\ref{fig:g-2mu-tanb}(a)}
\end{center}
\clearpage

~
\vfill
\begin{center}
\makebox[0cm]{
\def\epsfsize#1#2{\EPSSCALE#1}
\epsfbox{\EPSDIR 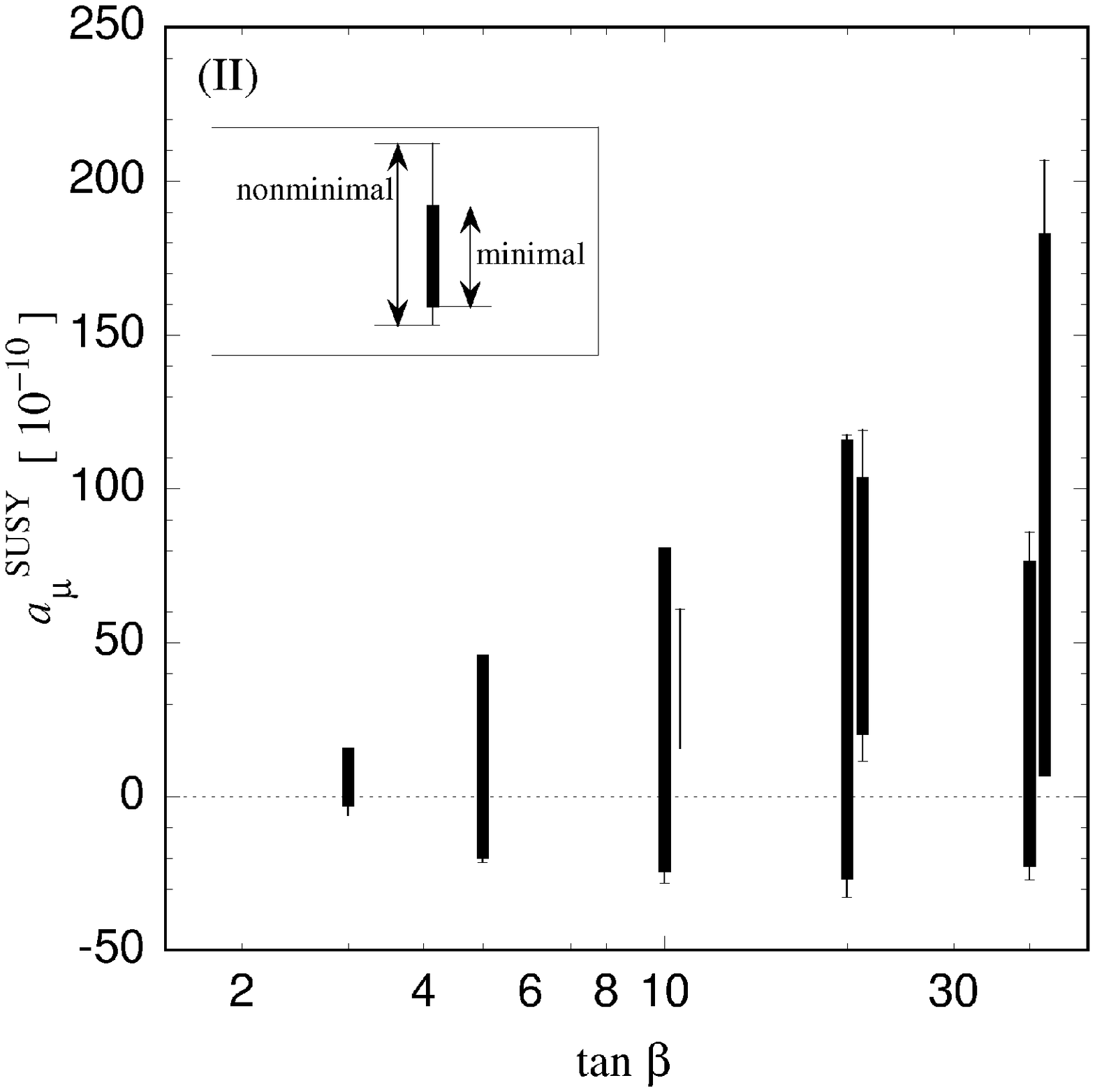}
}
\vfill
{\Large\bf Fig.~\ref{fig:g-2mu-tanb}(b)}
\end{center}
\clearpage

\end{document}